\documentclass[reprint, superscriptaddress, amsmath, amssymb, aps, prl, floatfix, longbibliography
]{revtex4-2}
\pdfoutput=1
\usepackage{CJK}
\usepackage{amsmath}
\usepackage{array}
\usepackage{dsfont}
\usepackage{mathtools}
\usepackage{graphicx}
\usepackage{dcolumn}
\usepackage{bm}
\usepackage[pdfencoding=auto]{hyperref}
\usepackage{physics}
\usepackage{empheq}
\usepackage{color}
\usepackage{multirow}
\usepackage{afterpage}
\usepackage{mathrsfs}
\usepackage{tikz}
\usepackage{filecontents}
\CJKnospace

\usepackage{xr}
\usetikzlibrary{shapes}
\usetikzlibrary{tikzmark,calc}

\begin{document}
\begin{CJK*}{UTF8}{gbsn}
\title{Towards a global phase diagram of Ce-based dipolar-octupolar pyrochlore magnets under magnetic fields
}

\author{Zhengbang Zhou ({\CJKfamily{gbsn}周政邦)}}
\email{zhengbang.zhou@mail.utoronto.ca}
\affiliation{%
 Department of Physics, University of Toronto, Toronto, Ontario M5S 1A7, Canada
}%
\author{Yong Baek Kim}%
\email{ybkim@physics.utoronto.ca}
\affiliation{%
 Department of Physics, University of Toronto, Toronto, Ontario M5S 1A7, Canada
}%

\date{\today}
\begin{abstract}
Recent experiments have established a strong case for Ce$_2$(Zr, Sn, Hf)$_2$O$_7$ to host $\pi$-flux quantum spin ice (QSI). However, an irrefutable conclusion still requires strong, multifaceted evidence. In dipolar-octupolar (DO) compounds, external magnetic fields only strongly couple with the dipolar component $\tau_z$ along its local z-axis in contrast to octupolar components $\tau^{x,y}$. This gives rise to the unique ways magnetic fields interact with the system and, in turn, provides us with a variety of tuning knobs to generate comprehensive experimental results. In this work, we focus on magnetic fields along the (110), (111), and (001) directions and present a plethora of remarkable experimental signatures to probe the underlying physics of $\pi$-flux QSI using gauge mean field theory (GMFT) and Monte Carlo simulations. In particular, we present unique signatures in magnetic field-dependent phase diagrams, equal-time and dynamical structure factors, and magnetostriction. 
\end{abstract} 
\maketitle
\end{CJK*}

\begin{figure*}
    \centering
    \includegraphics[width=0.9\linewidth]{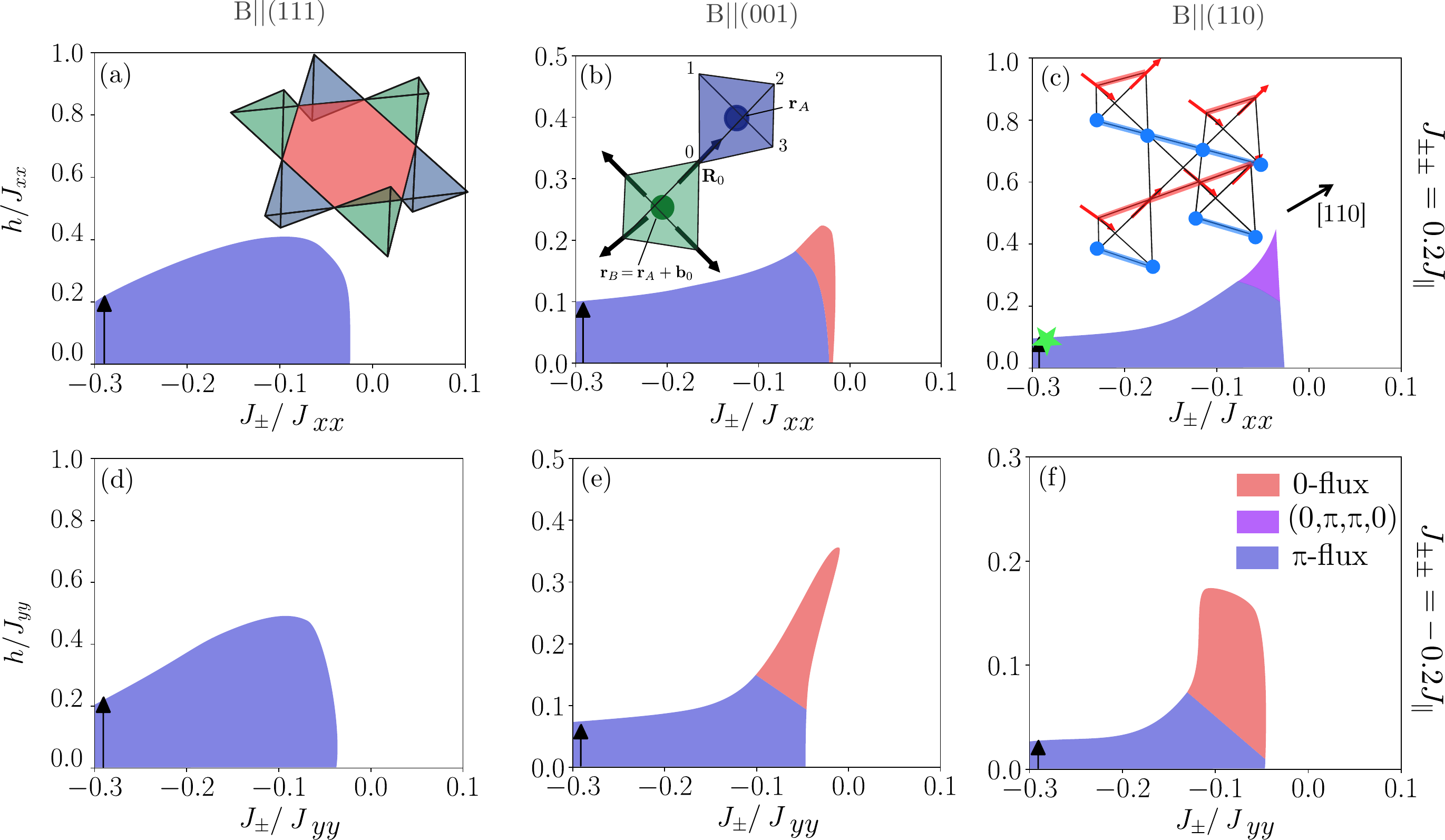}
    \caption{Phase diagrams containing dipolar parameter set $(J_{xx},J_{yy},J_{zz},J_{xz})=(0.063, 0.062, 0.011,0)$meV (a-c) and the octupolar parameter set $(J_{xx},J_{yy},J_{zz},J_{xz})=(0.062, 0.063, 0.011,0)$meV  (d-f) for Ce$_2$Zr$_2$O$_7$ denoted by the black arrow under field directions (111) (a,d), (001) (b,e), and (110) (c,f) by looking at the parameter spaces with $J_{\pm\pm}=0.2J_\parallel$ (a-c), $J_{\pm\pm}=-0.2J_\parallel$ (d-f), respectively. Blue denotes $\pi$-flux phase; red denotes 0-flux phase; and purple denotes the staggered $(0,\pi,\pi,0)$ phase. The green star in (c) denotes the parameter set for which the SSSF is computed. The insets: (a) an example hexagonal plaquette for which the flux thread through; (b) shows a pyrochlore parent unit cell denoting the pyrochlore coordinates $\mathbf{R}_\mu$ and the sublattice-indexed coordinate systems $\mathbf{r}_\alpha$ (see Sec.~\ref{I-sec:Appendix_DO_compound} in Ref.~\cite{supplementary} for definitions); (c) an example of the coupled $\alpha$ chain in red and the decoupled $\beta$ chain in blue under a (110) field.}
    \label{fig:CZO_phase_diagram}
\end{figure*}

\textit{Introduction}.--- Quantum spin liquids (QSLs) have garnered much theoretical attention due to their rich physics of hosting long-range entanglement (LRE) and fractionalized excitations~\cite{wen2002quantum, wen2004quantum, wen2013topological, balents2010spin, savary2016quantumspinliquids, knolle2019field}. However, experimental realization of such a state has remained a challenge to this day~\cite{bramwell2020thehistory, gingras2014quantum}. Recently, exciting experimental development has been made in Ce-based dipolar-octupolar (DO) materials, Ce$_2$(Zr, Sn, Hf)$_2$O$_7$~\cite{gaudet2019quantum, gao2019experimental, smith2022case, gao2022magnetic, smith2023dipole, smith2023quantum, Beare2023muSR, gao2024emergent, sibille2015candidate, yahne2022dipolar, sibille2020quantum, poree2025evidence, poree2024dipolar, bhardwaj2024thermodynamics, smith2025twopeakheatcapacityaccounts}, suggesting the possibility of hosting $\pi$-flux quantum spin ice (QSI). This is a three-dimensional quantum spin liquid with
a compact $U(1)$ emergent gauge structure, whose average flux threaded through hexagonal plaquettes formed by tetrahedron edges (see the inset Fig.~\ref{fig:CZO_phase_diagram}(a)) is $\pi$~\cite{hermele2004pyrochlore, ross2011quantum, banerjee2008unusual, shannon2012quantum, kato2015numerical, gingras2014quantum, udagawa2021spin, chern2019magnetic, castelnovo2012spin, balents2010spin, benton2012seeing, rau2019frustrated, savary2021quantum, lee2012generic, benton2018quantum, taillefumier2017competing, huang2018dynamics, bhardwaj2022sleuthing, bhardwaj2024thermodynamics,smith2025twopeakheatcapacityaccounts}. Heat capacity and muon spin relaxation measurements on these materials showed no sign of long-range order or spin freezing down to the lowest accessible experimental temperature~\cite{gaudet2019quantum,gao2019experimental,smith2022case,gao2022magnetic,smith2023quantum, Beare2023muSR, poree2022crystalfield, sibille2015candidate, poree2024dipolar}. Best-fitting microscopic exchange parameters from various measurements place these materials in the region of $\pi$-flux QSI~\cite{bhardwaj2022sleuthing, smith2022case, smith2023quantum, poree2024dipolar, bhardwaj2024thermodynamics, tang2013short, bhardwaj2022sleuthing, bhardwaj2024thermodynamics}. Furthermore, in Ce$_2$Zr$_2$O$_7$, the momentum-resolved energy-integrated dynamical spin structure factors obtained with neutron scattering~\cite{gaudet2019quantum, smith2022case, smith2023quantum} are in excellent qualitative agreement with theoretical predictions\cite{desrochers2023spectroscopic, Desrochers2024Finite, Hosoi2022Uncovering, chern2024pseudofermion}. More recently, a polarized neutron scattering experiment of the same material at low energy has also suggested the presence of the emergent photon modes unique to QSI states~\cite{gao2024emergent} where they also observed a $T^3$ dependence in the heat capacity.
Furthermore, the results of high-resolution backscattering neutron spectroscopy~\cite{poree2025evidence} results on Ce$_2$Sn$_2$O$_7$ have highlighted the presence of multiple peaks of decreasing intensity that was recently proposed as a characteristic signature of spinon excitations in QSI $\pi$-flux QSI~\cite{desrochers2023spectroscopic,desrochers2023symmetry, Desrochers2024Finite}. 

Despite these encouraging results, an unequivocal identification of QSI requires multifaceted evidence. In this regard, recent experiments~\cite{smith2023quantum} have shown that magnetic fields are powerful tuning knobs for generating a comprehensive experimental profile of $\pi$-flux QSI. More specifically, because Ce$_2$(Zr, Sn, Hf)$_2$O$_7$ are dipolar-octupolar compounds whose underlying pseudospin-1/2 operators $\tau^{x,y}$($\tau^z$) are octupolar(dipolar), magnetic fields couple strongly only with the dipolar pseudospin component $\tau^z$~\cite{rau2019frustrated, huang2014quantum, patri2020distinguishing} aligned along the local z-axis $\hat{\mathbf{z}}_\mu$ defined in Table~\ref{I-tab: Local basis} of Ref.~\cite{supplementary}. Here $\mu\in\{0,1,2,3\}$ is the pyrochlore sublattice index (see inset in Fig.~\ref{fig:CZO_phase_diagram}(b)). As such, given a magnetic field direction, the Zeeman coupling strength varies uniquely across the four tetrahedron sites, thereby giving us a rich parameter space to capture distinctive signatures of $\pi$-flux QSI~\cite{zhou2024magnetic}. In particular, in the case of $(110)$ direction, two sites (sites 0 and 3) are coupled to the magnetic field, and the other two (sites 1 and 2) are completely decoupled, forming systems of coupled ($\alpha$) and decoupled ($\beta$) one-dimensional chains, as shown in the inset in Fig.~\ref{fig:CZO_phase_diagram}(c). Another direction of interest is the $(111)$ field, where the much more strongly coupled sites (site 0) form a sparse triangle, and the others form a kagome lattice. Finally, in the case of $(001)$ field, all four sites are uniformly coupled along the field direction. Evidently, applications of magnetic fields can lower the space group symmetry and, in doing so, may give us access to unique signatures of the underlying $\pi$-flux QSI.

In this letter, we highlight important magnetic field responses of $\pi$-flux QSI candidates Ce$_2$(Zr, Sn, Hf)$_2$O$_7$ using the full XYZ model with the proposed microscopic exchange parameters. In particular, we first characterize the magnetic phase diagrams under various field directions for Ce$_2$(Zr, Hf)$_2$O$_7$. We highlight the critical field strength under a (110) field as a potential experimental marker for distinguishing the multipolar nature of Ce$_2$Zr$_2$O$_7$. After which, we compute, within the QSI phase, the distinctive signatures of static (equal-time) spin structure factor (SSSF) and dynamical spin structure factor (DSSF) for comparisons with neutron scattering data. We also show that the non-spin-flip channel accessible in polarized neutron scattering experiments is a distinctive signature to distinguish between 0-flux and $\pi$-flux phases. Finally, we showcase magnetostriction as a powerful probe to differentiate between degenerate parameter sets from fitting, helping us to discern the multipolar nature of the corresponding QSI phases under the context of Ce$_2$(Zr, Sn, Hf)$_2$O$_7$. 

\begin{figure*}
    \centering
    \includegraphics[width=0.95\linewidth]{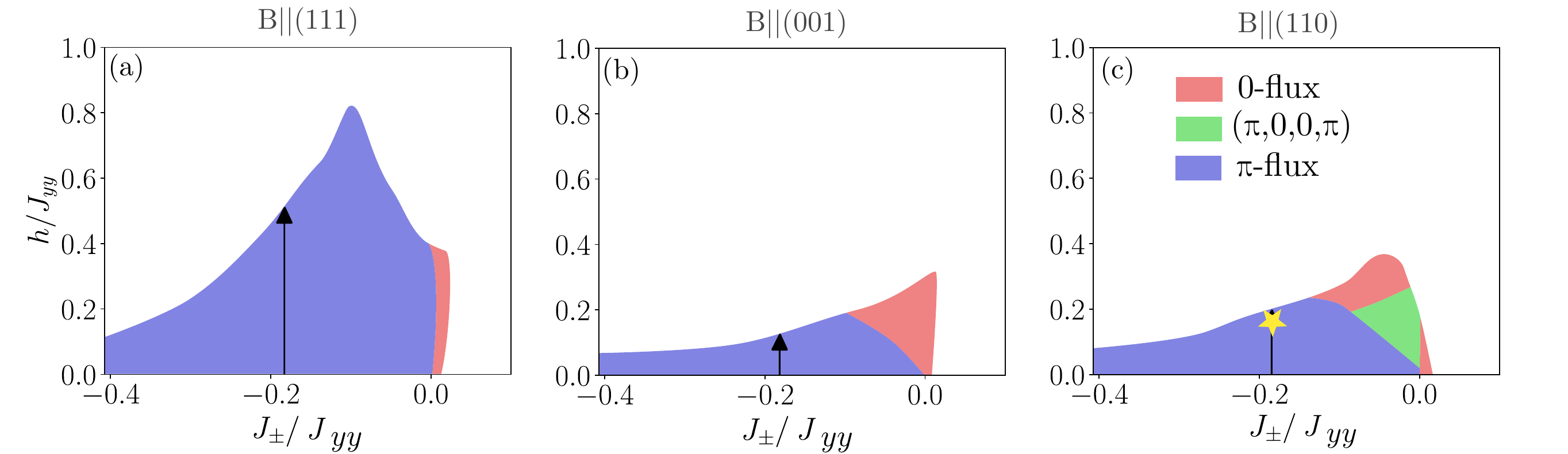}
    \caption{Phase Diagrams containing octupolar Ce$_2$Hf$_2$O$_7$ parameter $(J_{xx},J_{yy},J_{zz},J_{xz})=(0.020, 0.047, 0.013,-0.008)$meV denoted by the black arrows by picking the parameter space where $J_{\pm\pm}=-0.09J_{yy}$, $\theta=-0.58$ under magnetic field direction (111) (a), (001) (b), and (110) (c). Blue denotes $\pi$-flux phase; red denotes 0-flux phase; and green denotes the staggered $(\pi,0,0,\pi)$ phase. The yellow star denotes the parameter set in which we compute the SSSF with.}
    \label{fig:pyrochlore lattice}
\end{figure*}

\textit{Model}.--- The most general Hamiltonian for DO compounds coupled to external magnetic fields is the XYZ model, which, at the nearest neighbour level, takes the form of:
\begin{align}
    &\mathcal{H} = \sum_{\langle \mathbf{R}_\mu,\mathbf{R}_\nu \rangle} \left[J_{xx} S_{\mathbf{R}_\mu}^x S_{\mathbf{R}_\nu}^x +J_{yy} S_{\mathbf{R}_\mu}^y S_{\mathbf{R}_\nu}^y + J_{zz} S_{\mathbf{R}_\mu}^z S_{\mathbf{R}_\nu}^z \right]  \nonumber \\
    &- \mu_B \sum_{\mathbf{R}_\mu} \left[ (\mathbf{B} \cdot \hat{\mathbf{z}}_{\mu}) \left(g_{zz}\sin\theta S^x_{\mathbf{R}_\mu}  +  g_{zz}\cos\theta  S^z_{\mathbf{R}_\mu}\right)\right]\label{eq:H_full}.
\end{align}
Here $\mathbf{R}_\mu$ refers to the coordinate of the site within the sublattice-indexed pyrochlore coordinates (SIPC) defined in Eq.\eqref{I-eq:SIPC_SIDC} of Ref.~\cite{supplementary}. The $S^\alpha$ components are relate to pseudospin $\tau^\alpha$ via a rotation along the local $y$-axis by an angle $\theta$, where $\tau^y = S^y$; $\tau^x = \cos(\theta)S^x - \sin(\theta)S^z$; $\tau^z = \sin(\theta)S^x + \cos(\theta)S^z$; $\tan(2\theta)=\frac{2J_{xz}}{J_{xx}-J_{zz}}$. The land\'e g-factors of $\tau^{x,y}$ $g_{xx}$, $g_{yy}\sim0$ are calculated to be weak due to their octupolar nature~\cite{benton2020ground,huang2014quantum}. As such, we only consider the Zeeman term with $\tau^z$.
For the following discussion, without any loss of generalities, let us denote the dominant exchange coupling as $J_{\alpha\alpha} = J_\parallel$, where $J_{\alpha\alpha}=\max(J_{xx}, J_{yy}, J_{zz})$; the transverse coupling $J_\pm=-\frac{J_{\beta\beta}+J_{\gamma\gamma}}{4}$; $J_{\pm\pm}=\frac{J_{\beta\beta}-J_{\gamma\gamma}}{4}$ where $\{J_{\alpha\alpha}, J_{\beta\beta}, J_{\gamma\gamma}\}$ is some cyclic permutation of the Ising exchange interactions $\{J_{xx}, J_{yy}, J_{zz}\}$ depending on which is the dominant exchange parameter.

Experimentally relevant Ce-based compounds have exchange parameters predicted to be far away from the Ising limit ($J_\pm \ll J_{\parallel}$)~\cite{smith2023dipole, smith2022case, smith2023quantum,yahne2024dipolar, poree2024dipolar,sibille2018experimental,poree2025evidence}. As such, a non-perturbative approach is warranted for a faithful investigation. In particular, GMFT has been shown to be a powerful formalism in its ability to predict experimentally verified signatures in the $\pi$-flux QSI regime~\cite{wen2002quantum,liu2019competing,liu2021symmetric,schneider2022projective,chern2021theoretical, chern2017fermionic,chern2017quantum, desrochers2023symmetry, savary2016quantumspinliquids,savary2021quantum,desrochers2022competing,desrochers2023spectroscopic,Desrochers2024Finite} where other numerical techniques, such as quantum Monte Carlo (QMC), cannot be applied due to the sign problem~\cite{savary2016quantumspinliquids,gingras2014quantum, shannon2012quantum, shannon2021quantum, kato2015numerical, benton2020ground, huang2018dynamics}. This formalism maps the Hamiltonian in Eq.~\ref{eq:H_full} to a compact $U(1)$ gauge theory whereby the dominant pseudospin component $S^\alpha \sim E$ and $S^\pm = S^\beta \pm i S^\gamma \sim \Phi^\dagger e^{\pm i A}\Phi$~\cite{benton2012seeing, hermele2004pyrochlore, banerjee2008unusual, huang2018dynamics, pace2021emergent, Morampudi2020Spectroscopy}, where $A$ is the gauge field, $E$ is the corresponding electric field, and $\Phi^\dagger$, $\Phi$ are the creation annhilation operators for elementary excitations dubbed spinons. For the detailed GMFT formalism, please see Sec.~\ref{I-sec:Appendix_GMFT} in Ref.~\cite{supplementary}. 

\textit{Phase Diagram}.--- Before considering magnetic field responses, we first compute the phase diagrams to characterize these Ce-based compounds under finite magnetic fields of various directions. Since the exchange parameters are obtained via fitting with experimental data, there is often more than one set of acceptable parameters. To avoid ambiguity, for the rest of the investigation, we investigate both the dipolar-dominant Ce$_2$Zr$_2$O$_7$ parameters $(J_{xx},J_{yy},J_{zz},J_{xz})=(0.063, 0.062, 0.011,0)$meV~\cite{smith2023quantum} as well as the octupolar-dominant Ce$_2$Zr$_2$O$_7$ parameters $(J_{xx},J_{yy},J_{zz},J_{xz})=(0.062, 0.063, 0.011,0)$meV~\cite{smith2023quantum}. On the other hand for Ce$_2$Hf$_2$O$_7$, due to reasons discussed in Sec.~\ref{I-sec:Appendix_GMFT} of Ref.~\cite{supplementary}, we should expect GMFT formalism to be accurate only for the octupolar-dominant fitting parameters $(J_{xx},J_{yy},J_{zz},J_{xz})=(0.020, 0.047, 0.013,-0.008)$meV~\cite{poree2024dipolar}. For investigations into the dipolar-dominant case, we require techniques beyond the scope of this work. 

We, therefore, choose parameter spaces that contain the above parameter sets to compute the phase diagrams. Namely, we pick $J_{\pm\pm}=0.2J_\parallel$, $\theta=0$ for dipolar-dominant Ce$_2$Zr$_2$O$_7$, $J_{\pm\pm}=-0.2J_\parallel$, $\theta=0$ for octupolar-dominant Ce$_2$Zr$_2$O$_7$, and $J_{\pm\pm}=-0.09J_\parallel$, $\theta=-0.58$ for octupolar-dominant Ce$_2$Hf$_2$O$_7$. Then, we construct the phase diagrams as a function of $J_{\pm}$ and magnetic field $h$.

\begin{figure*}
    \centering
    \includegraphics[width=0.95\linewidth]{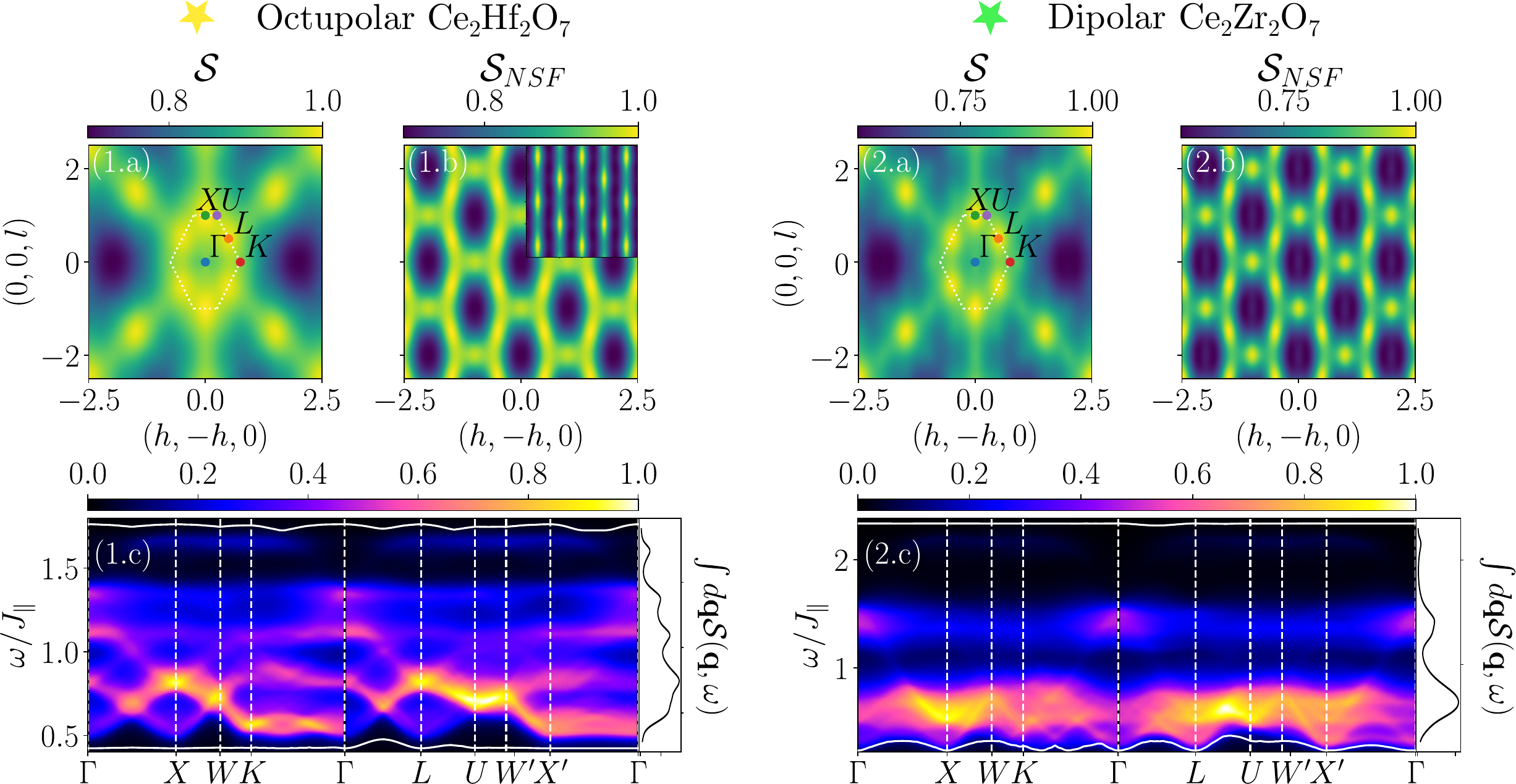}
    \caption{Static spin structure factor (1.a-1.b, 2.a-2.b) and dynamical spin structure factor (1.c, 2.c) with the proposed parameters of octupolar Ce$_2$Hf$_2$O$_7$ and dipolar Ce$_2$Zr$_2$O$_7$ under a (110) magnetic field with field strength $h=0.15J_{\parallel}$ (1.a-1.c) and $h=0.08J_{\parallel}$ (2.a-2.c). The inset in (1.b) shows $\mathcal{S}_{\text{NSF}}$ of an energetically less favourable 0-flux QSI under the octupolar Ce$_2$Hf$_2$O$_7$ parameters. Here, the DSSF is calculated along a path going through high symmetry points where $X=(1,0,0)$, $W=(1,-1/2,0)$, $K=(3/4,-3/4,0)$, $L=(1/2,1/2,1/2)$, $U=(1/4, 1/4, 1)$, $W'=(0,1/2,1)$ The side panels in (1.c) and (2.c) show the momentum integrated DSSF $\int d\mathbf{q}\mathcal{S}(\mathbf{q},\omega)$.} 
    \label{fig:CHO_SSF}
\end{figure*}

As shown in Fig.~\ref{fig:CZO_phase_diagram}, for Ce$_2$Zr$_2$O$_7$, where $J_\pm\approx -0.3J_{\parallel}$, GMFT predicts a rapidly vanishing $\pi$-flux QSI phase with small critical field strength $h_c$. Amongst which, we note that the $h_c$ under a (110) field for the dipolar case is $0.1J_\parallel$ which is an order of magnitude larger than that of the octupolar case where $h_c\approx0.02J_\parallel$. We note this remarkable qualitative difference can potentially serve as a way to distinguish the underlying multipolar nature of the $\pi$-flux QSI phase for Ce$_2$Zr$_2$O$_7$~\cite{smith2022case,smith2023quantum}. 

Due to the narrow region where QSI survives under magnetic fields for Ce$_2$Zr$_2$O$_7$, experiments are more often done outside of the QSI phase. We thereby provide classical phase diagrams accordingly at higher magnetic fields using Monte Carlo (MC) simulations in Sec.~\ref{I-sec:Appendix_phase_diagram} of Ref.~\cite{supplementary}. We see that, at higher field strength, Ce$_2$Zr$_2$O$_7$ is predicted to be in the classical spin liquid (CSL) regime before eventually crossing over into the polarized paramagnetic (PM) phase at $h\approx 3J_{\parallel}$ under a (111) field, $h\approx 2J_{\parallel}$ under a (001) field, and $h\approx 5J_{\parallel}$ under a (110) field.

On the other hand, due to the small $J_\pm$ and $J_{\pm\pm}$ for Ce$_2$Hf$_2$O$_7$, the predicted critical fields are quite significant as shown in Fig.~\ref{fig:pyrochlore lattice}: $h_c\approx0.12J_{\parallel}$ for $\mathbf{B}\parallel(001)$, $h_c\approx0.2J_{\parallel}$ for $\mathbf{B}\parallel(110)$, and $h_c\approx0.53J_{\parallel}$ for $\mathbf{B}\parallel(111)$. 
This provides a sizable region in the parameter space to study the magnetic field responses of $\pi$-flux QSI. We also observe the possibility of stabilizing a novel staggered flux phase dubbed $(\pi,0,0,\pi)$ under a (110) field near the Ising limit, which is a phase where half of the plaquettes are zero flux and the other half, $\pi$ flux~\cite{zhou2024magnetic}. 

\textit{Spin Structure Factors}.--- Now, we can compute SSSF and DSSF at experimentally meaningful parameters for Ce$_2$(Zr, Hf)$_2$O$_7$. Since both $\tau^x$ and $\tau^y$ have underlying octupolar magnetic charge distribution, $g_{xx}$, $g_{yy}\sim0$ and neutron scattering only probes correlations between pseudospin component $\tau^z$,~\cite{huang2014quantum, rau2019frustrated}. As such, we can compute the neutron scattering structure factor using: 
\begin{align}
\frac{\partial^2 \sigma }{\partial\Omega\partial\omega} &\propto \mathcal{S}(\mathbf{q},\omega)=\\
&\sum_{\mu\nu}\left(\hat{\mathbf{z}}_\mu \cdot \hat{\mathbf{z}}_\nu - \frac{(\hat{\mathbf{z}}_\mu\cdot \mathbf{q})(\hat{\mathbf{z}}_\nu\cdot \mathbf{q})}{|\mathbf{q}|^2} \right)  \mathcal{S}_{\mathrm{LF},\mu\nu}^{zz} (\mathbf{q},\omega) .
\end{align} Here $\mathcal{S}_{\mathrm{LF},\mu\nu}^{\alpha\beta} (\mathbf{q},\omega)$ is the dynamical spin correlations in the local sublattice-dependent frame 
$\mathcal{S}^{\alpha\beta}_{\mathrm{LF}, \mu\nu}(\mathbf{q},\omega)  = \sum_{\mathbf{R}_\mu, \mathbf{R}_\nu} e^{i\mathbf{q}(\mathbf{R}_\mu-\mathbf{R}_\nu)} \int dt e^{i\omega t}  \langle \tau^\alpha_{\mathbf{R}_\mu}(t)\tau^\beta_{\mathbf{R}_\nu}(0)\rangle.
$

We first highlight the SSSF signatures of octupolar-dominant $\pi$-flux QSI in Ce$_2$Hf$_2$O$_7$, defined as $\mathcal{S}(\mathbf{q})=\int d\omega \mathcal{S}(\mathbf{q},\omega)$, specifically under a (110) magnetic field as shown in Fig.~\ref{fig:CHO_SSF}(1.a) (see other directions in Sec.~\ref{I-sec:Appendix_SSSF} of Ref.~\cite{supplementary}). Here, we observe a similar snowflake pattern seen in $\pi$-flux QSI under zero field with pronounced intensities at the $L$ and $X$ points in the first Brillouin zone. On the other hand, the DSSF signature in Fig.~\ref{fig:CHO_SSF}(1.c) shows that the 3-peak structure, which is indicative of $\pi$-flux QSI under zero field, splits into 5 peaks by the (110) field. This, combined with the DSSF signature under other field directions, gives us a unique signature of $\pi$-flux QSI. 

One remark here is that the non-spin-flip (NSF) channel $\mathcal{S}_{\text{NSF}}$ under a (110) field can serve as an effective tool to probe the underlying physics of $\pi$-flux QSI. $\mathcal{S}_{\text{NSF}}$ (where neutrons are polarized perpendicular to the scattering plane) is defined by $\mathcal{S}_{\text{NSF}}(\mathbf{q},\omega)=\sum_{\mu\nu}\left(\hat{\mathbf{z}}_\mu \cdot \mathbf{\hat{v}} \right)\left( \hat{\mathbf{z}}_\nu \cdot \mathbf{\hat{v}} \right) \mathcal{S}_{\mathrm{LF},\mu\nu}^{zz} (\mathbf{q},\omega)$, where $\mathbf{\hat{v}}$ is the direction of neutron polarization. Due to the coupling with the local $z$-axis, when neutrons are polarized along the (110) direction, $\mathcal{S}_{\text{NSF}}(\mathbf{q},\omega)$ only contains correlations between sites 0 and 3 which constitute the $\alpha$ chains. As such, the NSF channel effectively separates the contribution between the coupled $\alpha$ and the uncoupled $\beta$ chains and allows direct observation of the former which are aligned perpendicular to the $(h,-h,l)$ plane. Therefore, $\mathcal{S}_{\text{NSF}}$ shown in Fig.~\ref{fig:CHO_SSF}(1.b) encodes strongly the information of the inter-$\alpha$-chain correlations. Here, the strong intensities along the Brillouin zone boundaries have been previously identified as characteristic of $\pi$-flux QSI~\cite{desrochers2023spectroscopic}. Moreover, we propose a further possibility of the NSF channel as a robust probe to distinguish between 0-flux and $\pi$-flux QSI under finite magnetic fields. As shown in the inset in Fig.~\ref{fig:CHO_SSF}(1.b.), where we investigate the energetically less favourable 0-flux QSI under the same parameters, $\mathcal{S}_{\text{NSF}}$ here is distinctively opposite to that of the $\pi$-flux QSI despite their SSSF signatures being qualitatively similar (see Sec.~\ref{I-sec:Appendix_SSSF} in Ref.~\cite{supplementary}). 

On the other hand, one important signature of $\pi$-flux QSI in Ce$_2$Zr$_2$O$_7$ can also be sleuthed out by looking at the NSF SSSF signature. Here we compute the SSSF along with the NSF signature of dipolar-dominant Ce$_2$Zr$_2$O$_7$. As shown in Fig.~\ref{fig:CHO_SSF}(2.b), $\mathcal{S}_{\text{NSF}}$ has modulated intensities along the Brillouin zone boundaries with minima at the $U$ points. This is due to a much stronger inter-$\alpha$-chain correlation along the $(00l)$ direction supplemented by a much larger $|J_\pm|$~\cite{zhou2024magnetic} unique to Ce$_2$Zr$_2$O$_7$ compared to the octupolar Ce$_2$Hf$_2$O$_7$ case in Fig.~\ref{fig:CHO_SSF}(1.b). 

\textit{Magnetostriction}.--- As previously discussed, there are often several acceptable fitting parameters for $\pi$-flux QSI candidates Ce$_2$(Zr, Sn, Hf)$_2$O$_7$~\cite{smith2023dipole, smith2022case, smith2023quantum,yahne2024dipolar, poree2024dipolar,sibille2018experimental,poree2025evidence}. To discern between the different sets of exchange parameters, we study magnetostriction signatures which have previously been shown to have distinctive behaviours between octupolar and dipolar 0-flux QSI~\cite{patri2020distinguishing,patri2020theory}. Namely, we consider the length change along (111), (110), and (001) ($\hat{\mathbf{p}}$ direction) when we apply a magnetic field in the (111), (110), and (001) ($\hat{\mathbf{h}}$ direction), denoted by $L^{\hat{\mathbf{h}}}_{\hat{\mathbf{p}}}$.  Here, the magnetostriction signatures are calculated from classical Monte Carlo simulations using magnetostriction parameters specified in Sec.~\ref{I-sec:Appendix_magneto} of Ref.~\cite{supplementary} at temperature $T=10^{-4}J_\parallel$.

In the case of Ce$_2$Zr$_2$O$_7$, the magnetostriction signatures between two parameter sets are qualitatively identical across all magnetic field directions~\cite{supplementary}. 

\begin{figure}
    \centering
    \includegraphics[width=\linewidth]{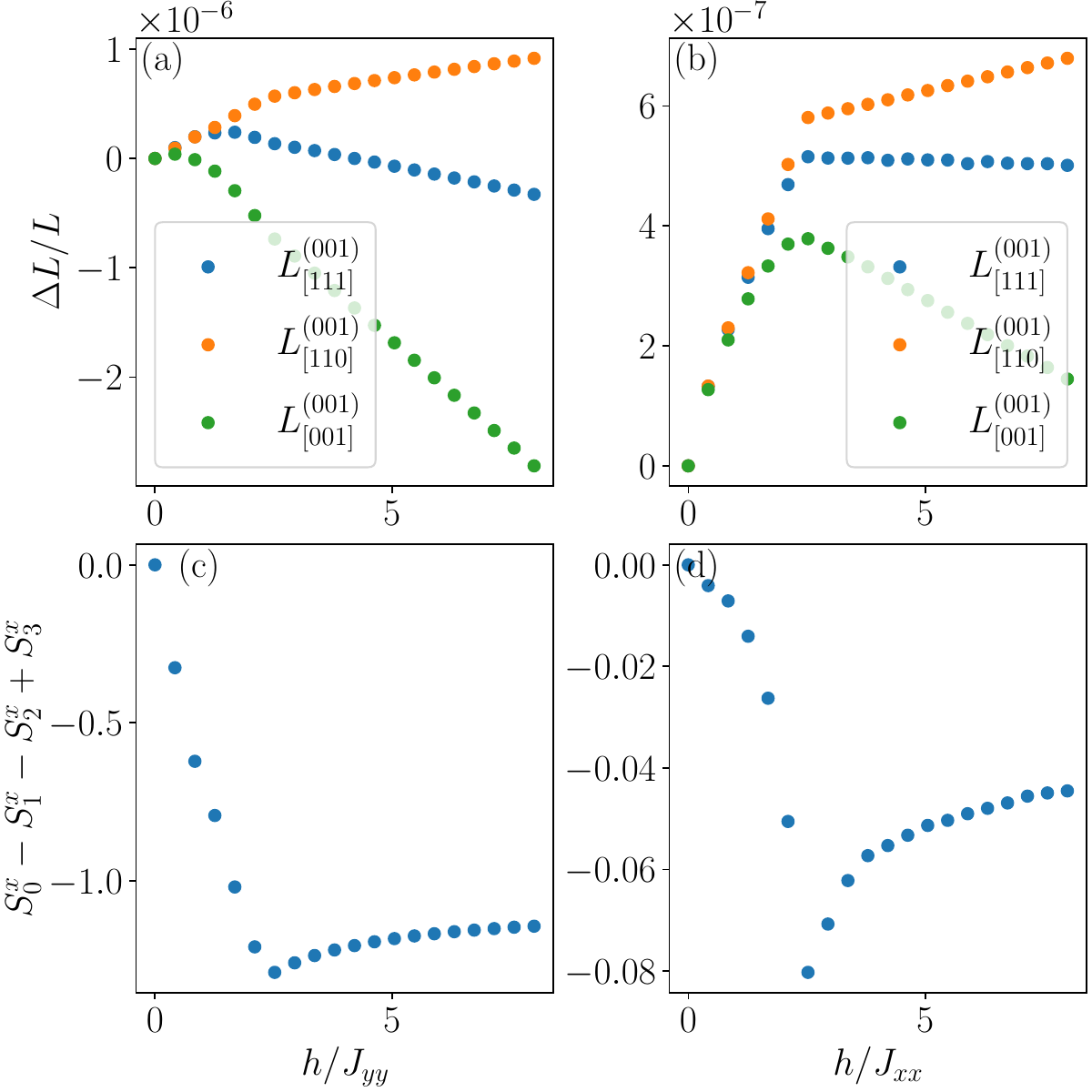}
    \caption{Magnetostriction signatures (a,b) $L^{(001)}_{[111]}$, $L^{(001)}_{[110]}$, $L^{(001)}_{[001]}$ and $S_{PM}^{x}= S_0^x-S_1^x-S_2^x+S_3^x$ (c,d) under the application of a (001) magnetic field for (a,c) octupolar parameter set and (b,d) dipolar parameter set of Ce$_2$Hf$_2$O$_7$.}
    \label{fig:magnetostriction}
\end{figure}

On the other hand, for Ce$_2$Hf$_2$O$_7$, where we select amongst the proposed parameters the octupolar-dominant QSI with $(J_{xx},J_{yy},J_{zz},J_{xz})=(0.020, 0.047, 0.013, -0.008)$meV and the dipolar-dominant QSI with $(J_{xx},J_{yy},J_{zz},J_{xz})=(0.046, 0.022, 0.011, -0.001)$meV. As shown in Fig.~\ref{fig:magnetostriction}(a-b) magnetostriction provides strikingly different signatures for the octupolar-dominant versus the dipolar-dominant case under a (001) field (see other field directions in Ref.~\cite{supplementary}). This qualitative difference is mainly due to the different underlying pseudospin configurations within each respective parameter set. For example,  $L^{(001)}_{[001]} \sim A h S_{PM}^{x} + B h S_{PM}^{z}$, where $S_{PM}^\alpha := S_0^\alpha-S_1^\alpha-S_2^\alpha+S_3^\alpha$ and $S_\mu^\alpha:=\langle S_{\mathbf{R}_\mu}^\alpha\rangle$. Here $A$ and $B$ are some constants depending on the magnetostriction coupling constants $g_{1}$ to $g_{10}$ (see Eq.~\eqref{I-eq:L001_001} in Ref.~\cite{supplementary}) and $h$ is the field strength. We find that as we increase the magnetic field, the behaviours of $S_{PM}^z$ between the dipolar and octupolar-dominant case are the same, whereas for $S_{PM}^x$ they are completely different, as shown in Fig.~\ref{fig:magnetostriction}(c-d). This is mainly because $\theta=-0.58$ for the octupolar-dominant case which is much larger than that of the dipolar-dominant case where $\theta=-0.03$. As such, the Zeeman term in Eq.~\eqref{eq:H_full} has a much stronger coupling with $S^x$ of the octupolar-dominant case, resulting in a much larger $|S_{PM}^x|$. We note that even though magnetostriction coupling constants are not known, within the same set of parameters, due to the drastically different behaviour of $S_{PM}^x$, the magnetostriction profile of dipolar versus octupolar-dominant regimes would always be qualitatively different, serving as a robust tool to distinguish between them. We further comment on the universality of this behaviour in Sec.\ref{I-sec:Appendix_magneto} of Ref.~\cite{supplementary}.

Furthermore, it is worth noting that the multipolar nature of Ce$_2$Sn$_2$O$_7$, whose possibility of hosting $\pi$-flux QSI is debated between two sets of proposed parameters~\cite{poree2025evidence,yahne2024dipolar}, can also be distinguished by looking at the magnetostriction signatures, as shown in Fig.~\ref{I-fig:CSO Magnetostriction}. 

\textit{Discussion}.--- In this letter, we used a combination of GMFT and MC to explore the magnetic responses of DO compounds. We computed the QSI magnetic phase diagrams for Ce$_2$(Zr, Hf)$_2$O$_7$ to categorize the regions in parameters space where $\pi$-flux QSI is stable. In doing so, we found that the critical field strengths under a (110) field are an order of magnitude apart between dipolar and octupolar parameter sets of Ce$_2$Zr$_2$O$_7$. We also presented predictions for SSSF and DSSF of Ce$_2$(Zr, Hf)$_2$O$_7$. Here, beyond the distinctive marker seen in the SSSF, we show that the NSF channel under a (110) field can serve as an effective tool in distinguishing 0-flux and $\pi$-flux QSI. Finally, in further aiding distinguishing degenerate fitting parameter sets, we present predictions of magnetostriction signatures of Ce$_2$(Zr, Sn, Hf)$_2$O$_7$ under various magnetic field directions. In short, we provide definite predictions for a series of powerful experimental results for elucidating the nature of the spin liquid states in the QSI candidates Ce$_2$(Zr, Sn, Hf)$_2$O$_7$.
\begin{acknowledgments}
We thank F\'elix Desrochers, Minoru Yamashita, and Adarsh S. Patri for the many helpful discussions. This work was supported by the Natural Science and Engineering Council of Canada (NSERC) Discovery Grant No. RGPIN-2023-03296 and the Center for Quantum Materials at the University of Toronto. Computations were performed on
the Niagara cluster, which SciNet hosts in partnership
with the Digital Research Alliance of Canada. Z.Z. is supported by the Ontario Graduate Scholarship (OGS). 
\end{acknowledgments}

\bibliography{ref}

\makeatletter\@input{xx.tex}\makeatother
\end{document}


\title{Supplemental Material for \linebreak
``Towards a global phase diagram of Ce-based dipolar-octupolar pyrochlore magnets under magnetic fields''}

\author{Zhengbang Zhou}
\email{zhengbang.zhou@mail.utoronto.ca}
\affiliation{%
 Department of Physics, University of Toronto, Toronto, Ontario M5S 1A7, Canada
}%
\author{Yong Baek Kim}%
\email{ybkim@physics.utoronto.ca}
\affiliation{%
 Department of Physics, University of Toronto, Toronto, Ontario M5S 1A7, Canada
}%

\date{\today}

\maketitle

\tableofcontents

\setcounter{secnumdepth}{3}
\setcounter{equation}{0}
\renewcommand{\theequation}{S\arabic{equation}}
\setcounter{table}{0}
\renewcommand{\thetable}{S\arabic{table}}
\setcounter{figure}{0}
\renewcommand{\thefigure}{S\arabic{figure}}

\newpage
\section{Dipolar-Octupolar Pyrochlore \label{sec:Appendix_DO_compound}}

Sublattice-indexed pyrochlore coordinates (SIPC), $\mathbf{R}_\mu$, and sublattice indexed diamond coordinates (SIDC), $\mathbf{r}_\alpha$ are used in the main text to label the pyrochlore sites and the parent diamond lattice sites, respectively. They are related through:
\begin{align}
     \mathbf{R}_{\mu}  &= r_1 \mathbf{e}_1 + r_2 \mathbf{e}_2 + r_3 \mathbf{e}_3 - \eta_{\alpha}\mathbf{b}_0/2 + \eta_{\alpha}\mathbf{b}_{\mu}/2  \notag\\
     &= \mathbf{r}_{\alpha} + \eta_{\alpha}\mathbf{b}_{\mu}/2, \label{eq:SIPC_SIDC}
\end{align}
where $\eta_\alpha=1 \text{ if } \alpha=A$ \text{or} $\eta_\alpha=-1 \text{ if } \alpha=B$. $\mathbf{b}_\mu$ are vectors connecting the center of a down-pointing tetrahedron to the centers of its nearest up-pointing tetrahedrons
\begin{subequations}
\begin{align}
    \mathbf{b}_0 &= -\frac{1}{4}(1,1,1)\\
    \mathbf{b}_1 &= \frac{1}{4}(-1,1,1)\\
    \mathbf{b}_2 &= \frac{1}{4}(1,-1,1)\\
    \mathbf{b}_3 &= \frac{1}{4}(1,1,-1), \label{eq:bmu}
\end{align}
\end{subequations}
and $\mathbf{e}_i$ are the lattice basis vectors
\begin{subequations}
\begin{align}
    \mathbf{e}_0 &= (0,0,0)\\
    \mathbf{e}_1 &= \frac{1}{2}(0,1,1)\\
    \mathbf{e}_2 &= \frac{1}{2}(1,0,1)\\
    \mathbf{e}_3 &= \frac{1}{2}(1,1,0).
\end{align}
\end{subequations}
We have introduced $\mathbf{e}_0=\mathbf{0}$ for convenience.

Spins are defined in sublattice-dependent local frames whose basis vectors are given in table \ref{tab: Local basis}.
\begin{table}[!ht]
\caption{\label{tab: Local basis}%
Local sublattice basis vectors
}
\begin{ruledtabular}
\begin{tabular}{ccccc}
$\mu$ & 0 & 1  & 2  & 3 \\
\hline
$\hat{x}_{\mu}$ & $\frac{1}{\sqrt{6}}\left(-2,1,1\right)$ & $\frac{-1}{\sqrt{6}}\left(2,1,1\right)$  & $\frac{1}{\sqrt{6}}\left(2,1,-1\right)$  & $\frac{1}{\sqrt{6}}\left(2,-1,1\right)$    \\
$\hat{y}_{\mu}$ & $\frac{1}{\sqrt{2}}\left(0,-1,1\right)$  & $\frac{1}{\sqrt{2}}\left(0,1,-1\right)$  & $\frac{-1}{\sqrt{2}}\left(0,1,1\right)$ & $\frac{1}{\sqrt{2}}\left(0,1,1\right)$  \\[2mm]
$\hat{z}_{\mu}$ & $\frac{1}{\sqrt{3}}\left(1,1,1\right)$ & $\frac{-1}{\sqrt{3}}\left(-1,1,1\right)$  & $\frac{-1}{\sqrt{3}}\left(1,-1,1\right)$  & $\frac{-1}{\sqrt{3}}\left(1,1,-1\right)$   \\[2mm]
\end{tabular}
\end{ruledtabular}
\end{table}

Furthermore, we find that generators for the pyrochlore space group:
\begin{subequations}\label{eq:PyrochloreSGgen}
\begin{align}
T_i&: \mathbf{r}_\alpha \mapsto\left(r_1+\delta_{i, 1}, r_2+\delta_{i, 2}, r_3+\delta_{i, 3}\right)_\alpha \\
\bar{C}_6&: \mathbf{r}_\alpha \mapsto\left(-r_3,-r_1,-r_2\right)_{\pi(\alpha)} \\
S&: \mathbf{r}_\alpha \mapsto\left(-r_1,-r_2, r_1+r_2+r_3+\delta_{\alpha, A}\right)_{\pi(\alpha)}.
\end{align}
\end{subequations}

\section{Gauge Mean Field Theory\label{sec:Appendix_GMFT}}
We elaborate here on the gauge mean field theory (GMFT) formalism~\cite{savary2012coulombic, lee2012generic, savary2016quantumspinliquids, li2017symmetry, yao2020pyrochlore, desrochers2023symmetry}. For the sake of simplicity, let us first derive this with the assumption that $J_{yy}$ is dominant (i.e. $|J_{yy}|>|J_{xx}|,|J_{zz}|$, $J_\parallel=J_{yy}$). First, we introduce a slave matter field $Q\in \mathbb{Z}$ on the parent diamond lattice defined as $Q_{\mathbf{r}_\alpha} = \sum_{\mu\in \partial t_{\mathbf{r}_\alpha}} S_{\mathbf{R}_\mu}^y$
where $t_{\mathbf{r}_\alpha}$ denotes the tetrahedron centered at $\mathbf{r}_\alpha$ and $\partial t_{\mathbf{r}_\alpha}$ refers to the four pyrochlore sites forming its boundary. The index $\alpha\in\{A,B\}$ corresponds to up-pointing and down-pointing tetrahedrons. The associated charge raising and lowering operators can then be naturally defined using the conjugate variable $\varphi_{\mathbf{r}_\alpha}$ (i.e., $[\varphi_{\mathbf{r}_\alpha}, Q_{\mathbf{r}_\alpha'}]=i\delta_{\mathbf{r}_\alpha \mathbf{r}_\alpha'}$) to be $\Phi_{\mathbf{r}_\alpha}^\dagger = e^{i\varphi_{\mathbf{r}_\alpha}}$ and $\Phi_{\mathbf{r}_\alpha} = e^{-i\varphi_{\mathbf{r}_\alpha}}$. By construction, the length of this operator is $|\Phi_{\mathbf{r}_\alpha}^\dagger \Phi_{\mathbf{r}_\alpha}| = 1$. However, it is found that GMFT only produces more agreeable critical $J_\pm^c$ with Quantum Monte Carlo (QMC) results for 0-flux QSI if this constraint is relaxed to $\langle\Phi_{\mathbf{r}_\alpha}^\dagger \Phi_{\mathbf{r}_\alpha}\rangle = \kappa$, where $\kappa=2$~\cite{desrochers2023symmetry, desrochers2023spectroscopic}. We will enforce this constraint by introducing Lagrange multipliers $\lambda^\alpha$.

Next, the original pseudospin operators can be extended to act on the enlarged Hilbert space $\mathcal{H}=\mathcal{H}_Q \times \mathcal{H}_{spin}$ via $S_{\mathbf{R}_\mu}^{+} \rightarrow \Phi^\dagger_{\mathbf{r}_A} \left( \frac{1}{2}e^{iA_{\mathbf{r}_A, \mathbf{r}_A + \mathbf{b}_\mu}} \right)\Phi_{\mathbf{r}_A+\mathbf{b}_\mu}$ and $S_{\mathbf{R}_\mu}^{y} \rightarrow E_{\mathbf{r}_A, \mathbf{r}_A+\mathbf{b}_\mu}$, 
where $E_{\mathbf{r}_A, \mathbf{r}_A+\mathbf{b}_\mu}$ and $A_{\mathbf{r}_A, \mathbf{r}_A + \mathbf{b}_\mu}$ are canonical conjugate electric and gauge fields acting on the initial spin Hilbert space, $\mathbf{b}_\mu$ are vectors connecting A diamond sites to the nearest four B diamond sites defined in Eq.~\eqref{eq:bmu}, and $\mathbf{R}_\mu$ is the pyrochlore site connecting the up and down tetrahedron centered at $\mathbf{r}_A$ and $\mathbf{r}_A+\mathbf{b}_{\mu}$ respectively. We therefore arrive at the following Hamiltonian.
\begin{widetext}
\begin{align}\label{eq:H_GMFT_FULL}
\mathcal{H}= & \frac{J_{\parallel}}{2} \sum_{\mathbf{r}_\alpha} Q_{\mathbf{r}_\alpha}^2-\frac{J_{ \pm}}{4} \sum_{\mathbf{r}_\alpha} \sum_{\mu, \nu \neq \mu} \Phi_{\mathbf{r}_\alpha+\eta_\alpha \mathbf{b}_\mu}^{\dagger} \Phi_{\mathbf{r}_\alpha+\eta_\alpha \mathbf{b}_\nu} e^{i \eta_\alpha\left(A_{\mathbf{r}_\alpha, \mathbf{r}_\alpha+\eta_\alpha \mathbf{b}_\nu}-A_{\mathbf{r}_\alpha, \mathbf{r}_\alpha+\eta_\alpha \mathbf{b}_\mu}\right)} \nonumber \\
& +\frac{J_{ \pm \pm}}{8} \sum_{\mathbf{r}_\alpha} \sum_{\mu, \nu \neq \mu}\left(e^{i \eta_\alpha\left(A_{\mathbf{r}_\alpha, r_\alpha+\eta_\alpha \mathbf{b}_\nu}+A_{\mathbf{r}_\alpha, \mathbf{r}_\alpha+\eta_\alpha \mathbf{b}_\mu}\right)} \Phi_{\mathbf{r}_\alpha}^{\dagger} \Phi_{\mathbf{r}_\alpha}^{\dagger} \Phi_{\mathbf{r}_\alpha+\eta_\alpha \mathbf{b}_\mu} \Phi_{\mathbf{r}_\alpha+\eta_\alpha \mathbf{b}_\nu}+\text { h.c. }\right) \nonumber \\
&- \frac{h}{4} \sum_{\mathbf{r}_A, \mu} (\hat{\mathbf{n}} \cdot \hat{\mathbf{z}}_{\mu}) \left((\cos\theta -i\sin\theta)\Phi^\dagger_{\mathbf{r}_A} \Phi_{\mathbf{r}_A + \mathbf{b}_\mu} e^{iA_{\mathbf{r}_A, \mathbf{r}_A+\mathbf{b}_\mu}} + h.c.\right),
\end{align}
\end{widetext}
where $J_{\pm}=-(J_{xx}+J_{zz})/4$ and $J_{\pm\pm}=(J_{zz}-J_{xx})/4$.
To obtain a tractable theory, the gauge field $A$ is completely fixed by its mean field configuration $\bar{A}$ and we do not consider any gauge fluctuation. In doing so, we also integrate out the canonical electric field $E$. Furthermore, we perform mean-field decoupling on the quartic term in $J_{\pm\pm}$:
\begin{align}
\Phi_{\mathbf{r}_\alpha}^{\dagger} \Phi_{\mathbf{r}_\alpha}^{\dagger} \Phi_{\mathbf{r}_\alpha+\eta_\alpha \mathbf{b}_\mu} \Phi_{\mathbf{r}_\alpha+\eta_\alpha \mathbf{b}_\nu} &\rightarrow \langle \Phi_{\mathbf{r}_\alpha}^{\dagger} \Phi_{\mathbf{r}_\alpha}^{\dagger} \rangle \Phi_{\mathbf{r}_\alpha+\eta_\alpha \mathbf{b}_\mu} \Phi_{\mathbf{r}_\alpha+\eta_\alpha \mathbf{b}_\nu} + \Phi_{\mathbf{r}_\alpha}^{\dagger} \Phi_{\mathbf{r}_\alpha}^{\dagger} \langle \Phi_{\mathbf{r}_\alpha+\eta_\alpha \mathbf{b}_\mu} \Phi_{\mathbf{r}_\alpha+\eta_\alpha \mathbf{b}_\nu} \rangle \nonumber\\ &+ \langle \Phi_{\mathbf{r}_\alpha}^{\dagger}  \Phi_{\mathbf{r}_\alpha+\eta_\alpha \mathbf{b}_\mu} \rangle \Phi_{\mathbf{r}_\alpha}^{\dagger}\Phi_{\mathbf{r}_\alpha+\eta_\alpha \mathbf{b}_\nu}
\langle \Phi_{\mathbf{r}_\alpha}^{\dagger} \Phi_{\mathbf{r}_\alpha+\eta_\alpha \mathbf{b}_\nu}\rangle \Phi_{\mathbf{r}_\alpha}^{\dagger} \Phi_{\mathbf{r}_\alpha+\eta_\alpha \mathbf{b}_\mu} \nonumber\\ &+ 
\Phi_{\mathbf{r}_\alpha}^{\dagger} \langle \Phi_{\mathbf{r}_\alpha}^{\dagger} \Phi_{\mathbf{r}_\alpha+\eta_\alpha \mathbf{b}_\mu} \rangle \Phi_{\mathbf{r}_\alpha+\eta_\alpha \mathbf{b}_\nu} + \Phi_{\mathbf{r}_\alpha}^{\dagger} \langle \Phi_{\mathbf{r}_\alpha}^{\dagger}  \Phi_{\mathbf{r}_\alpha+\eta_\alpha \mathbf{b}_\nu}\rangle \Phi_{\mathbf{r}_\alpha+\eta_\alpha \mathbf{b}_\mu}.
\end{align}
Therefore, we introduce mean field parameters $\overline{\chi}_{0}= \langle \Phi_{\mathbf{r}_\alpha}^{\dagger} \Phi_{\mathbf{r}_\alpha}^{\dagger} \rangle$, $\chi_{\mu\nu}=\langle \Phi_{\mathbf{r}_\alpha+\eta_\alpha \mathbf{b}_\mu} \Phi_{\mathbf{r}_\alpha+\eta_\alpha \mathbf{b}_\nu} \rangle $, and $\xi_{\mathbf{r}_\alpha,\mathbf{r}_\alpha+\eta_\alpha\mathbf{b}_\mu}=\langle \Phi_{\mathbf{r}_\alpha}^{\dagger} \Phi_{\mathbf{r}_\alpha+\eta_\alpha \mathbf{b}_\mu} \rangle $ to obtain the following mean field Hamiltonian: 
\begin{widetext}
\begin{align}\label{eq:H_GMFT}
\mathcal{H}_{\mathrm{GMFT}} &=  \frac{J_{\parallel}}{2} \sum_{\mathbf{r}_\alpha} Q_{\mathbf{r}_\alpha}^2+\sum_{\mathbf{r}_\alpha} \lambda^\alpha\left(\Phi_{\mathbf{r}_\alpha}^{\dagger} \Phi_{\mathbf{r}_\alpha}-\kappa\right)-\frac{J_{ \pm}}{4} \sum_{\mathbf{r}_\alpha} \sum_{\mu, \nu \neq \mu} \Phi_{\mathbf{r}_\alpha+\eta_\alpha \mathbf{b}_\mu}^{\dagger} \Phi_{\mathbf{r}_\alpha+\eta_\alpha \mathbf{b}_\nu} e^{i \eta_\alpha\left(\bar{A}_{\mathbf{r}_\alpha, \mathbf{r}_\alpha+\eta_\alpha \mathbf{b}_\nu}-\bar{A}_{\mathbf{r}_\alpha, \mathbf{r}_\alpha+\eta_\alpha \mathbf{b}_\mu}\right)} \nonumber \\
& +\frac{J_{ \pm \pm}}{8} \sum_{\mathbf{r}_\alpha} \sum_{\mu, \nu \neq \mu}\left[e ^ { i \eta _ { \alpha } ( \bar{A} _ { \mathbf { r } _ { \alpha } , \mathbf { r } _ { \alpha } + \eta _ { \alpha } \mathbf { b } _ { \nu } } + \bar{A}_{ \mathbf { r } _ { \alpha } , \mathbf { r } _ { \alpha } + \eta _ { \alpha } \mathbf { b } _ { \mu } } ) } \left(\Phi_{\mathbf{r}_\alpha}^{\dagger} \Phi_{\mathbf{r}_\alpha}^{\dagger} \chi_{\mathbf{r}_\alpha+\eta_\alpha \mathbf{b}_\mu, \mathbf{r}_\alpha+\eta_\alpha \mathbf{b}_\nu}+\bar{\chi}_{\mathbf{r}_\alpha, \mathbf{r}_\alpha}^0 \Phi_{\mathbf{r}_\alpha+\eta_\alpha \mathbf{b}_\mu} \Phi_{\mathbf{r}_\alpha+\eta_\alpha \mathbf{b}_\nu}\right.\right. \nonumber \\
&\left.\left.+2 \Phi_{\mathbf{r}_\alpha}^{\dagger} \Phi_{\mathbf{r}_\alpha+\eta_\alpha \mathbf{b}_\mu} \xi_{\mathbf{r}_\alpha, \mathbf{r}_\alpha+\eta_\alpha \mathbf{b}_\nu}+2 \Phi_{\mathbf{r}_\alpha}^{\dagger} \Phi_{\mathbf{r}_\alpha+\eta_\alpha \mathbf{b}_\nu} \xi_{\mathbf{r}_\alpha, \mathbf{r}_\alpha+\eta_\alpha \mathbf{b}_\mu}\right)+ h.c.\right] \notag\\
&- \frac{h}{4} \sum_{\mathbf{r}_A, \mu} (\hat{\mathbf{n}} \cdot \hat{\mathbf{z}}_{\mu}) \left[(\cos\theta-i\sin\theta)\Phi^\dagger_{\mathbf{r}_A} \Phi_{\mathbf{r}_A + \mathbf{b}_\mu} e^{i\bar{A}_{\mathbf{r}_A, \mathbf{r}_A+\mathbf{b}_\mu}} + h.c.\right]
\end{align}
\end{widetext}

 Fitting parameter sets of Ce-based QSI candidates can also be that of dipolar natures with dominant $J_{xx}$ exchange parameters. In this case, we instead map $S^x\sim E$, $S^y=\frac{S^++S^-}{2}$, $S^z=\frac{S^+-S^-}{2i}$, where $S^\pm=\Phi^\dagger e^{\pm iA}\Phi$. This would give us the same Hamiltonian but with a different Zeeman term in Eq.\eqref{eq:H_GMFT_FULL} where we now have 
\begin{align}
    \mathcal{H}_{\text{Zeeman}}&=- \frac{h}{4} \sum_{\mathbf{r}_A, \mu} (\hat{\mathbf{n}} \cdot \hat{\mathbf{z}}_{\mu}) \left[\sin\theta E_{\mathbf{r}_A,\mathbf{r}_A+\mathbf{b}_\mu}-i\cos\theta\Phi^\dagger_{\mathbf{r}_A} \Phi_{\mathbf{r}_A + \mathbf{b}_\mu} e^{i\bar{A}_{\mathbf{r}_A, \mathbf{r}_A+\mathbf{b}_\mu}} + h.c.\right].
\end{align}
So now, instead of strictly coupled to spinons, magnetic fields can also couple to the electric field with coupling strength $\propto\sin\theta$. However, notice that in the GMFT formalism, the electric field is completely integrated out. As such, a part of the Zeeman term is completely neglected. Therefore, we should be careful when applying GMFT to dipolar QSI if $\theta\neq0$ where the coupling with the electric field is not zero, as the computed phase diagrams might not be completely accurate. In particular, we note that for the dipolar-dominant fitting of Ce$_2$Hf$_2$O$_7$ $(J_{xx},J_{yy},J_{zz},J_{xz})=(0.046,0.022,0.011,-0.001)$meV, $\theta=-0.0285\neq0$. As such, we refrained from showing the dipolar phase diagrams of Ce$_2$Hf$_2$O$_7$.

\begin{figure*}
    \centering
    \includegraphics[width=0.9\linewidth]{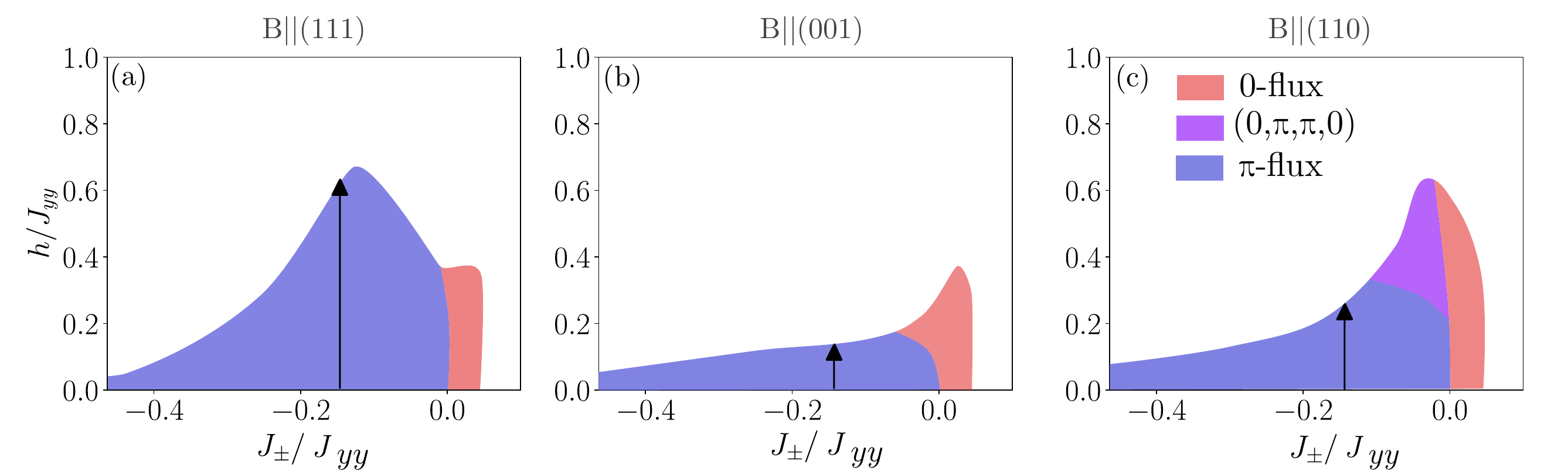}
    \caption{Phase Diagrams containing octupolar Ce$_2$Hf$_2$O$_7$ parameter $(J_{xx},J_{yy},J_{zz},J_{xz})=(0.011,0.044,0.016,-0.002)$meV denoted by the black arrows by picking the parameter space where $J_{\pm\pm}=0.04J_{yy}$, $\theta=0.34$ under magnetic field direction (111) (a), (001) (b), and (110) (c). Blue denotes $\pi$-flux phase; red denotes 0-flux phase; and purple denotes the staggered $(0,\pi,\pi,0)$ phase. The yellow star denotes the parameter set in which we compute the SSSF with.}
    \label{fig:CHO_phase_diagram_B}
\end{figure*}

Furthermore, for completeness, we also computed phase diagrams for another set of acceptable parameters $(J_{xx},J_{yy},J_{zz},J_{xz})=(0.011,0.044,0.016,-0.002)$meV~\cite{poree2024dipolar} as shown in Fig.~\ref{fig:CHO_phase_diagram_B}.

We note that Ce$_2$Zr$_2$O$_7$ has proposed fitting parameters with $\theta\sim0$~\cite{smith2023quantum}. As such, GMFT is still reasonably candid for both the dipolar and octupolar fitting parameters of Ce$_2$Zr$_2$O$_7$. The same can be said for one study of Ce$_2$Hf$_2$O$_7$~\cite{smith2025twopeakheatcapacityaccounts}, where the parameter sets are also considered to have $\theta\sim 0$, whereas another proposes a possible non-zero $\theta$ dipolar-dominant QSI~\cite{smith2025twopeakheatcapacityaccounts}.

\section{Supplementary Spin Structure Factors\label{sec:Appendix_SSSF}}

\begin{figure}
    \centering
    \includegraphics[width=\linewidth]{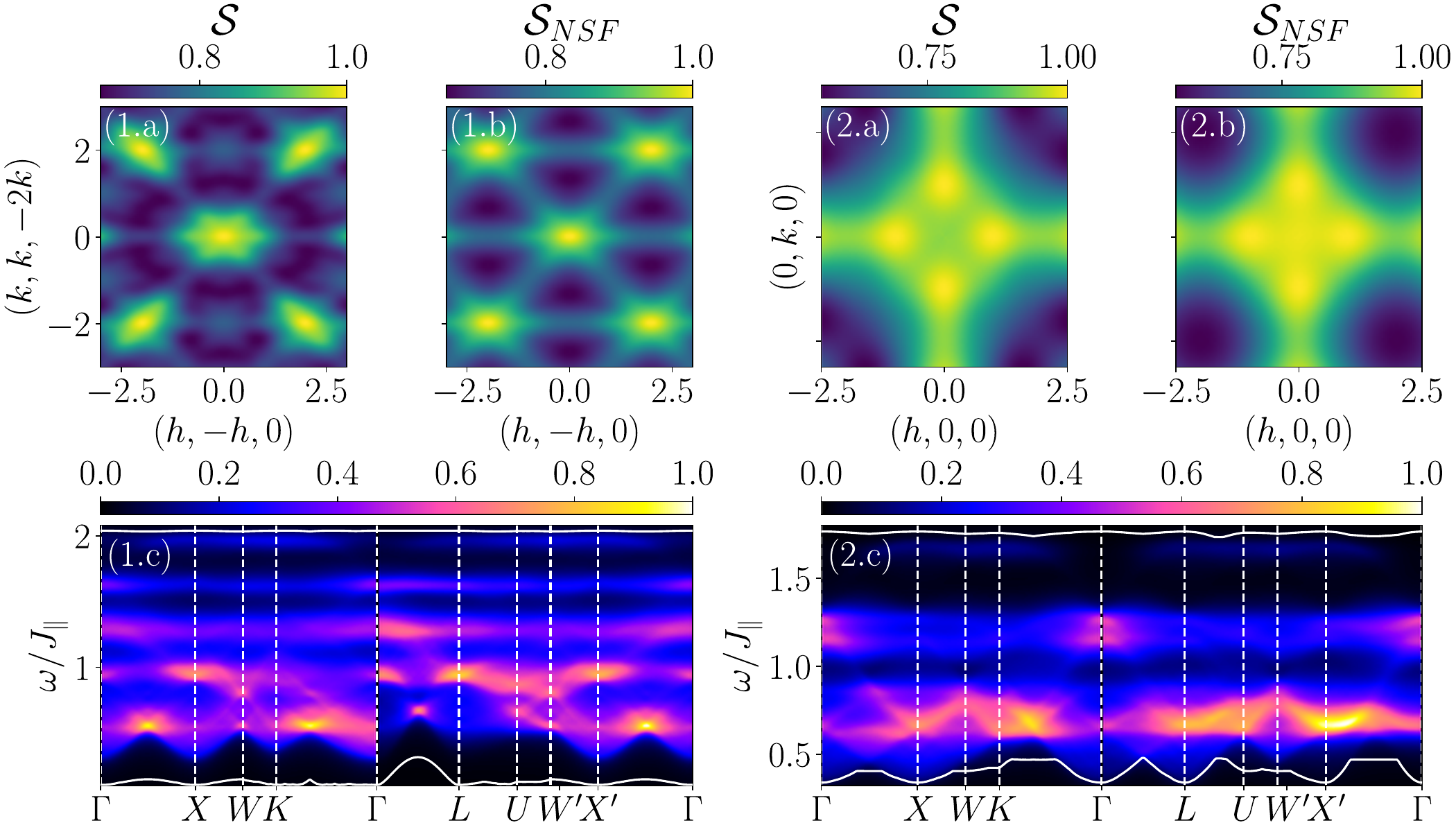}
    \caption{Static spin structure factor (1.a-1.b, 2.a-2.b) and dynamical spin structure factor (1.c, 2.c) with the proposed parameters of octupolar Ce$_2$Hf$_2$O$_7$ under a (111) magnetic field with field strength $h=0.4J_{\parallel}$ (1.a-1.c) and a (001) $h=0.1J_{\parallel}$ (2.a-2.c).}
    \label{fig:SSF_Supp_CHO}
\end{figure}

\begin{figure}
    \centering
    \includegraphics[width=\linewidth]{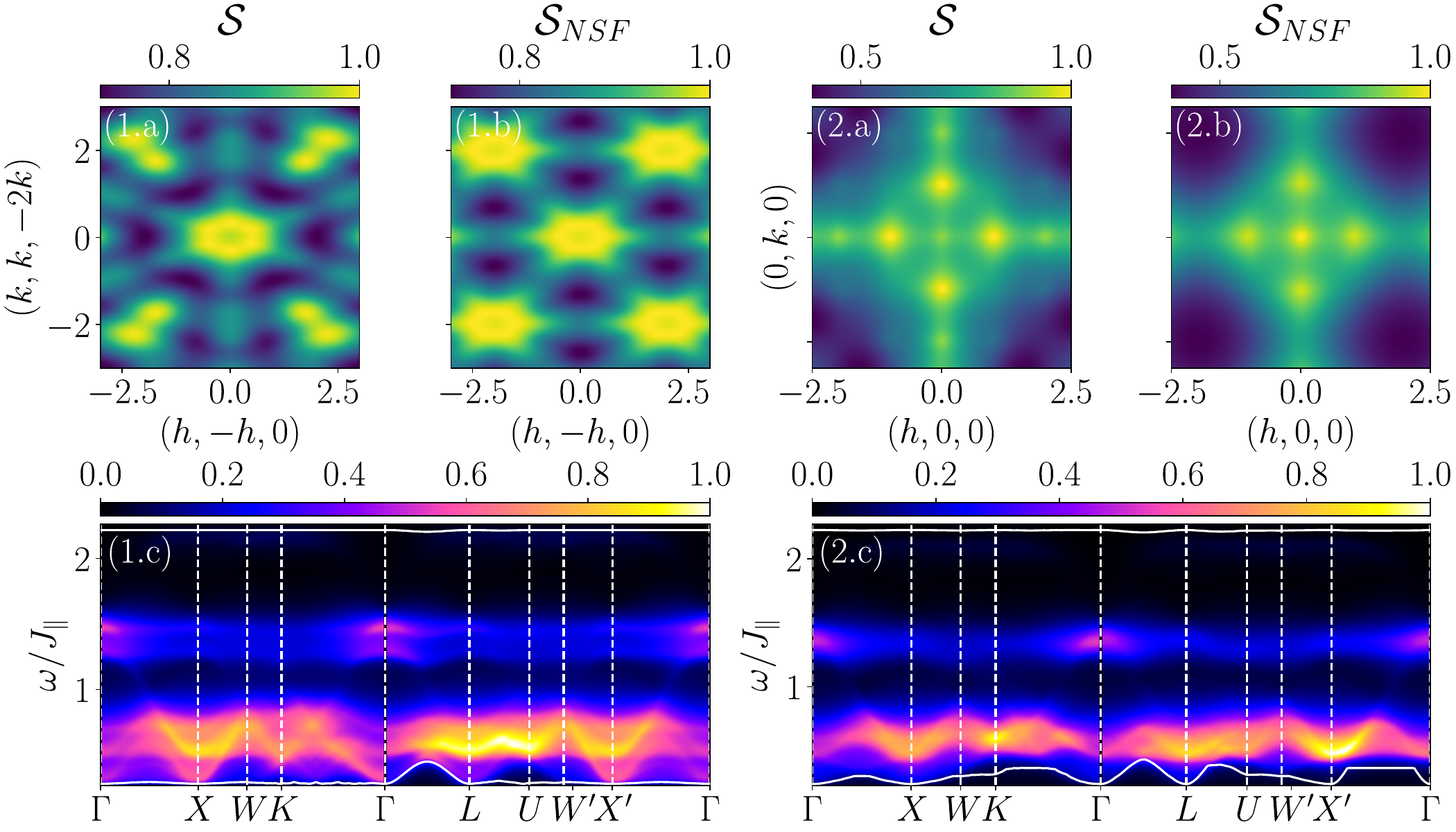}
    \caption{Static spin structure factor (1.a-1.b, 2.a-2.b) and dynamical spin structure factor (1.c, 2.c) with the proposed parameters of dipolar Ce$_2$Zr$_2$O$_7$ under a (111) magnetic field with field strength $h=0.1J_{\parallel}$ (1.a-1.c) and a (001) $h=0.05J_{\parallel}$ (2.a-2.c).}
    \label{fig:SSF_Supp_CZO}
\end{figure}

We present here some supplementary spin structure factors to elaborate on some details discussed in the main text. First of all, we present the SSSF and DSSF signatures of other field directions, (111) and (001), for the octupolar-dominant parameter set of Ce$_2$Hf$_2$O$_7$ in Fig.~\ref{fig:SSF_Supp_CHO} and Ce$_2$Zr$_2$O$_7$ in Fig.~\ref{fig:SSF_Supp_CZO} at various field strengths.

\begin{figure}
    \centering
    \includegraphics[width=0.5\linewidth]{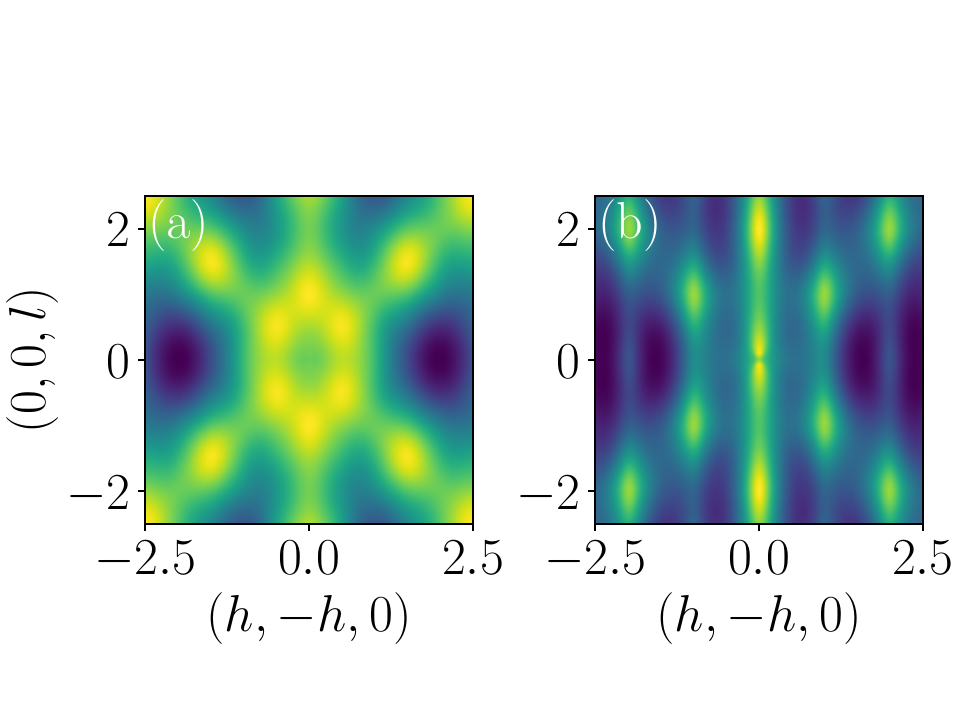}
    \caption{Static spin structure factor $\mathcal{S}$ with the proposed parameters of octupolar Ce$_2$Hf$_2$O$_7$ under a (110) magnetic field with field strength $h=0.15J_{\parallel}$ with the energetically favourable $\pi$-flux phase (a) and the less favourable $0$-flux phase (b).}
    \label{fig:SSSF_pi_flux_0_flux}
\end{figure}

We discussed in the main text the potential of the non-spin-flip channel as a distinctive marker to distinguish 0-flux and $\pi$-flux QSI. We want to supplement this argument by providing the equal time structure factor (SSSF) under the same parameter sets between the energetically favourable $\pi$-flux phase and the less favourable 0-flux phase in Fig.~\ref{fig:SSSF_pi_flux_0_flux}. On close inspection, strictly speaking, these SSSF signatures are different. Namely, the peak intensities of the 0-flux phase are not along the first Brillouin zone boundary like that of the $\pi$-flux phase. However, the overall similar snowflake pattern and the closeness in these peak positions might be hard to resolve under momentum-resolved neutron scattering data. We contrast this with the strikingly opposite signatures for the non-spin-flip channel discussed in the main text.

\section{Classical Simulations\label{sec:classical_sim}}
The classical simulations were done with Monte Carlo algorithm on a $4\times12\times12\times12$ pyrochlore lattice. We performed $10^7$ sweeps per temperature step for careful spin configuration convergence. After which, to obtain the DSSF, we then time evolve $\langle S^\alpha \rangle$ after simulated annealing by applying the classical equation of motion for SU(2) spin, which is the Landau-Liftshitz-Gilbert equation,
\begin{equation}
    \frac{\partial \mathbf{S}_{\mathbf{R}_\mu}}{\partial t} = - \mathbf{S}_{\mathbf{R}_\mu} \times  \frac{\partial \mathcal{H}}{\partial  \mathbf{S}_{\mathbf{R}_\mu} }.
\end{equation}

\subsection{Classical Phase Diagram \label{sec:Appendix_phase_diagram}}
\begin{figure*}
    \centering
    \includegraphics[width=\linewidth]{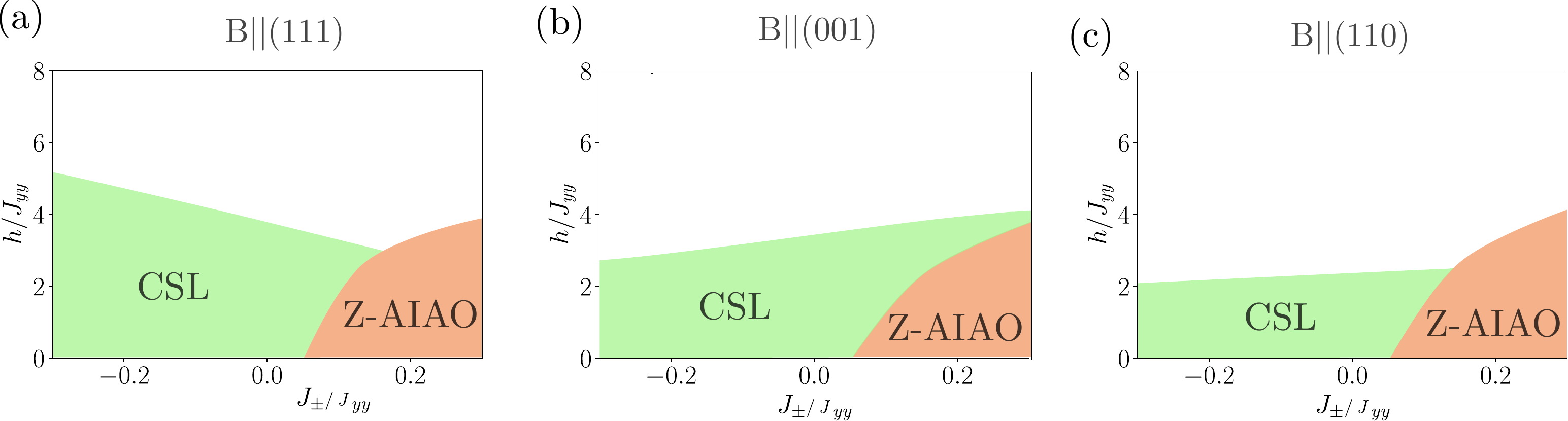}
    \caption{Classical phase diagram with $J_\pm=-0.2$, $\theta=0$ under magnetic field directions (a) (111), (b) (001), and (c) (110). Here green denotes the classical spin liquid phase (CSL); orange denotes the all-in-all-out in the $S^z$ basis (Z-AIAO).}
    \label{fig:classical phase}
\end{figure*}

We present the classical phase diagrams containing parameter sets for octupolar-dominant Ce$_2$Zr$_2$O$_7$ in Fig.~\ref{fig:classical phase} at temperature $T=10^{-4}|J_{\parallel}|^{-1}$. Since we are performing classical Monte Carlo simulations, the zero field ground states are that of classical spin ices (CSIs). As we increase the magnetic fields, the classical spin ice phases cross over into classical spin liquid (CSL) phases, where there still exists extensive ground state degeneracy. We refer to this phase as the CSL phase to make distinctions from the CSI phase since the residual entropy is not that of Pauling ice entropy and the underlying spin configurations are not exactly 2-in-2-out. We elaborate on this more in the discussions below. 

To determine the phase boundaries between the classical spin liquid phase and the paramagnetic phase, we compute the entropy up to $T=10^{-3}|J_{\parallel}|^{-1}$. Here, since we assume that $J_{yy}$ is dominant, at zero field, the system is in a degenerate manifold of 2-in-2-out spins aligned in the local y-axis (Y-2I2O). On the other hand, as previously discussed, the magnetic field only couples to $S^z$ degrees of freedom since $\theta=0$. Therefore, as we increase the magnetic field, the spins will gradually align with the local z-axis until we reach the paramagnetic spin configuration under each respective field direction. During this process, the spin degeneracy of Y-2I2O is not completely lifted. To see this, let us consider a simplified case where $J_{yy}\neq0$ and $J_{xx}=J_{zz}=0$ under a finite (110) field. At zero fields, one example of the y-components of the spins is $\{S^y_0,S^y_1,S^y_2,S^y_3\}:=\{1/2, 1/2, -1/2, -1/2\}$ where the expectation values of all other components are zero. Notice that here, any permutation of $\{S^y_\mu\}$ will yield the same energy, giving us a degenerate manifold of Y-2I2O spin configuration with a residual entropy of $\frac{1}{2}\ln \frac{3}{2}$. On the other hand, for a finite magnetic field, since the Zeeman term only couples with sites 0 and 3 with the same coupling strength albeit with opposite sign, $|\langle S^z_{0}\rangle| = |\langle S^z_{3}\rangle| \neq 0$. We see that one example of the y-components of the spins are $\{S^y_0,S^y_1,S^y_2,S^y_3\}:=\{\langle S^y_0\rangle, 1/2, -1/2, \langle S^y_3\rangle\}$, where $|\langle S^y_\mu\rangle|^2+|\langle S^z_\mu\rangle|^2=S^2=1/4$ for $\mu\in\{0,3\}$. As such, $|\langle S^y_0\rangle| = |\langle S^y_3\rangle|$. Therefore, we see that for a given spin configuration, swapping the spin components $S^y$ on sites 0 and 3 or on sites 1 and 2 are degenerate in energy, since the energy associated with the Zeeman term only depends on the z-components $\langle S^z_\mu\rangle$. We note that this residual entropy should be less than the Pauling ice entropy as we only have 4 degenerate configurations per tetrahedron. We can do the same for the other two directions. We find that, for the (001) direction, the residual entropy is approximately that of the Pauling entropy. On the other hand, we should expect the residual entropy of Kagome-ice for the (111) field~\cite{bojesen2017Quantum,fukazawa2002magnetic, sakakibara2003observation, cornelius2001shortrange}. Of course, the above derivation only is true only for $J_{xx}=J_{zz}=0$. Nevertheless, for $J_\pm\neq0$ and $J_{\pm\pm}\neq0$, we should expect a finite residual degeneracy for the CSL phases, unlike the polarized paramagnetic phases where the entropy is 0. As such, we can gain a conclusive phase boundary of the CSL phase by looking at the residual entropy.

On the other hand, we note that the phase whose spin configuration is the all-in-all-out in the local z-axis (Z-AIAO) is stable up to $h\sim 4J_{yy}$ before transitioning to the paramagnetic (PM) phase. Since the PM spin configuration also aligns in the local z-axis but strictly not AIAO, the Z-AIAO configuration directly competes with the PM phase, resulting in a first-order transition from the Z-AIAO phase to the PM phase. As such, delineations between the Z-AIAO and the PM phase can be found by looking at $\langle S^{z}_{\mathbf{R}_\mu}\rangle$. In the Z-AIAO phase, $\langle S^{z}_{\mathbf{R}_\mu}\rangle=1/2$, whereas in all the other PM configurations, $\langle S^{z}_{\mathbf{R}_\mu}\rangle<1/2$. 

\subsection{Magnetostrictions \label{sec:Appendix_magneto}}

\begin{figure*}
    \centering
    \includegraphics[width=\linewidth]{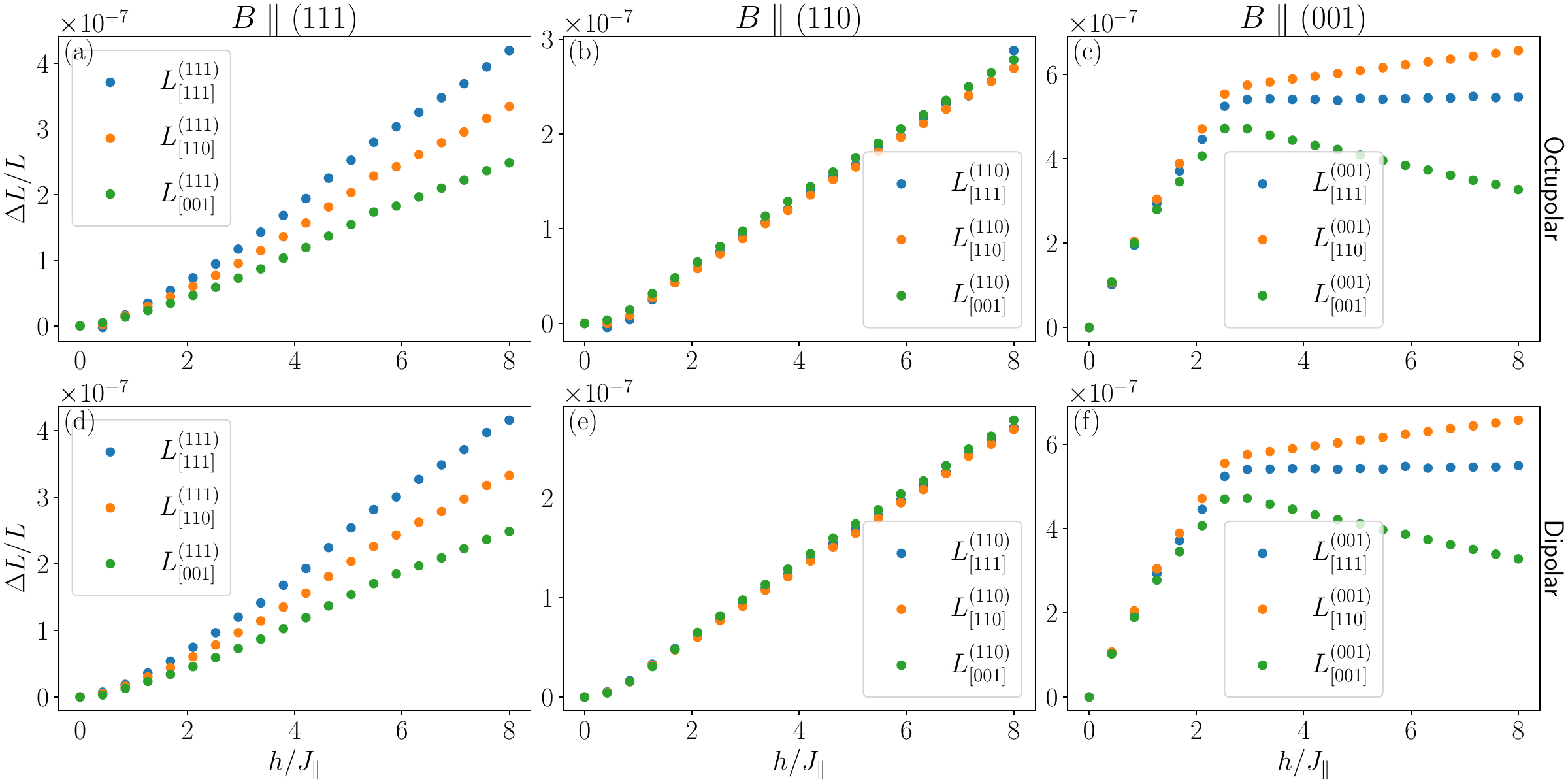}
    \caption{Magnetostriction for Ce$_2$Zr$_2$O$_7$ under octupolar-dominant fitting parameters-dominant $(J_{xx},J_{yy},J_{zz},J_{xz})=(0.062, 0.063, 0.011, 0)$meV (a-c) and the dipolar-dominant parameter set with $(J_{xx},J_{yy},J_{zz},J_{xz})=(0.063, 0.062, 0.011, 0)$meV under magnetic field direction (111) (a,d), (110) (b,e), and (001) (c,f).}
    \label{fig:CZO Magnetostriction}
\end{figure*}

\begin{figure*}
    \centering
    \includegraphics[width=\linewidth]{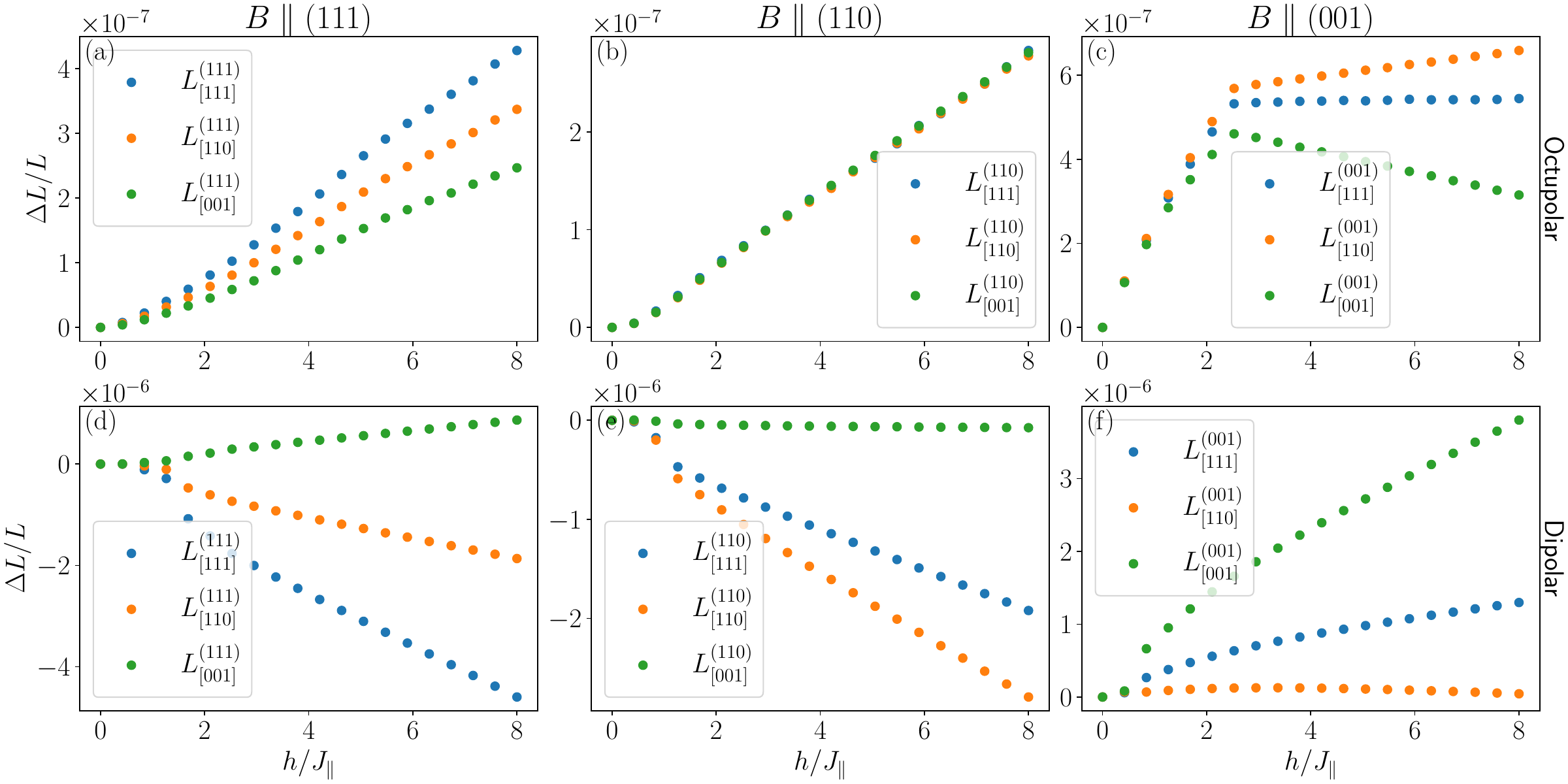}
    \caption{Magnetostriction for Ce$_2$Sn$_2$O$_7$ under octupolar-dominant fitting parameter $(J_{xx},J_{yy},J_{zz})=(0.0104, 0.048, 0.0104)$meV, $\theta=0$ (a-c) and the dipolar-dominant parameters with $(J_{xx},J_{yy},J_{zz})=(0.045, -0.001, -0.012)$meV, $\theta=0.19\pi$ under magnetic field direction (111) (a,d), (110) (b,e), and (001) (c,f).}
    \label{fig:CSO Magnetostriction}
\end{figure*}

\begin{figure*}
    \centering
    \includegraphics[width=\linewidth]{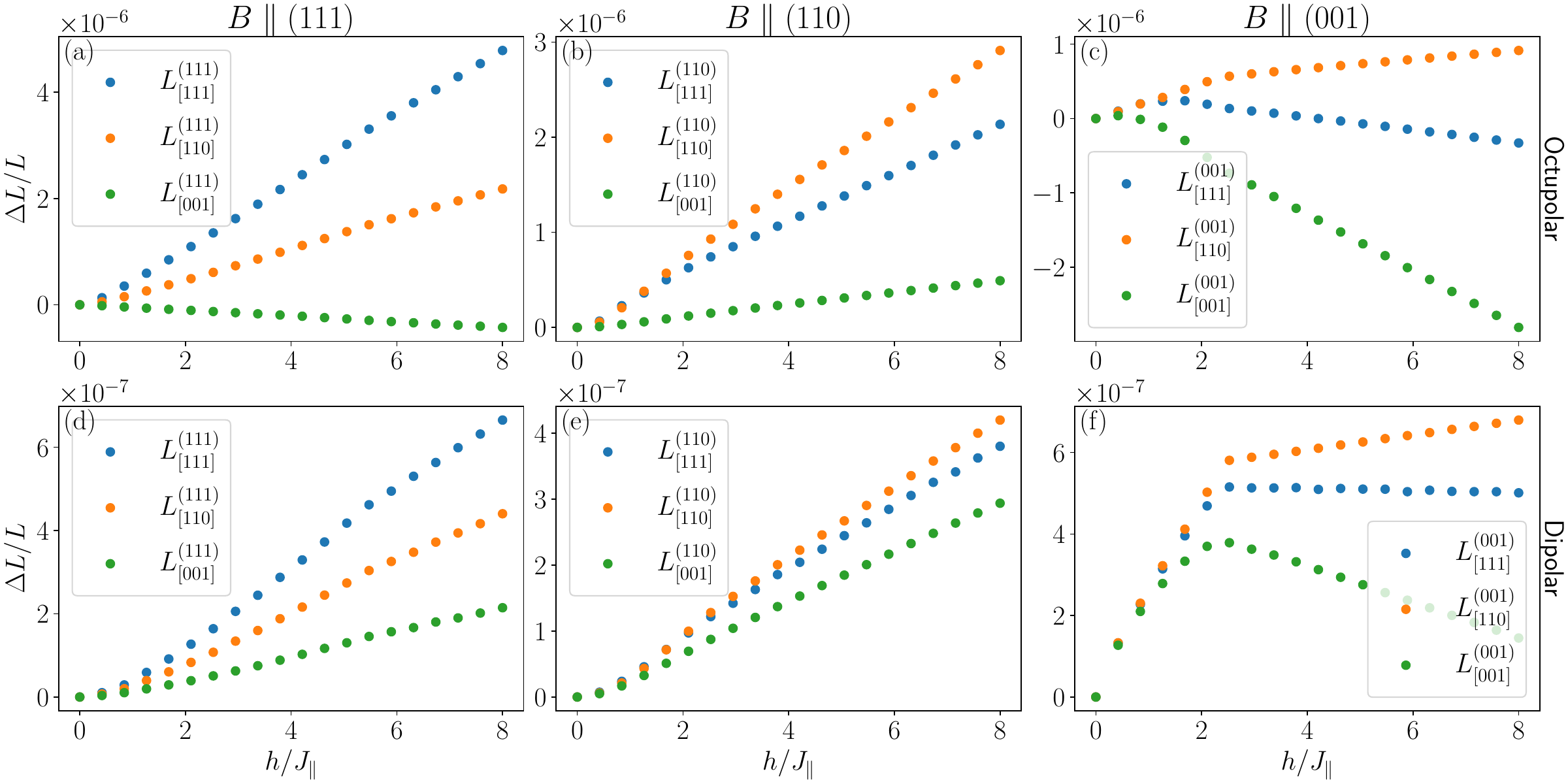}
    \caption{Magnetostriction for Ce$_2$Hf$_2$O$_7$ under octupolar-dominant
 fitting parameter $(J_{xx},J_{yy},J_{zz},J_{xz})=(0.020, 0.047, 0.013, -0.008)$meV (a-c) and the dipolar-dominant parameter set with $(J_{xx},J_{yy},J_{zz},J_{xz})=(0.046, 0.022, 0.011, -0.001)$meV (d-f) under magnetic field direction (111) (a,d), (110) (b,e), and (001) (c,f).}
    \label{fig:CHO Magnetostriction}
\end{figure*}

We present the magnetostriction calculation under magnetic field directions (111), (110), and (001) for Ce$_2$(Zr, Sn, Hf)$_2$O$_7$ in Fig.~\ref{fig:CZO Magnetostriction}, \ref{fig:CSO Magnetostriction}, \ref{fig:CHO Magnetostriction} respectively. 
The magnetostriction calculations follow the same derivation from Ref~\cite{patri2020distinguishing, patri2020theory}. To aid in a helpful discussion of physical understandings of the computed results, we hereby merely quote the relevant equation for calculating magnetostriction length change $L^{(001)}_{[001]}$ for discussions in the main text.

\begin{flalign}\label{eq:L001_001}
\left(\frac{\Delta L}{L}\right)_{[001]}^{(001)}= & \frac{1}{9 \sqrt{3} c_{33}} h\left[\left(2 g_3+g_4\right)\left(S_0^x-S_1^x-S_2^x+S_3^x\right)+\left(2 g_9+g_{10}\right)\left(S_0^z-S_1^z-S_2^z+S_3^z\right)\right] \nonumber\\
& +\frac{1}{3 \sqrt{3}\left(c_{11}-c_{22}\right)} h\left[\left(\sqrt{2} g_1+g_2\right)\left(S_0^x-S_1^x-S_2^x+S_3^x\right)+\left(\sqrt{2} g_7+g_8\right)\left(S_0^z-S_1^z-S_2^z+S_3^z\right)\right] .
\end{flalign}

Here, to evaluate the magnetostriction length change, we consider the following parameters: $g_1 = 4.7 \times 10^{-7}, g_2 =6.2\times 10^{-7}, g_3 = 8.0\times10^{-7}, g_4 =-3.06\times10^{-7}, g_5 = 0.176\times 10^{-7}, g_6 = -1.78\times 10^{-7}, g_7 = -0.26\times10^{-7}, g_8 =-0.51\times10^{-7}, g_9 = 0.78\times10^{-7}, g_{10} = 1.09\times10^{-7}$, $c_{33}=1$, $c_{11}-c_{22}=1$.

We compute the magnetostriction length change profiles between the two degenerate fitting parameters of Ce$_2$Zr$_2$O$_7$, $(J_{xx},J_{yy},J_{zz},J_{xz})=(0.062, 0.063, 0.011, 0)$meV and $(J_{xx},J_{yy},J_{zz},J_{xz})=(0.063, 0.062, 0.011, 0)$meV~\cite{smith2023quantum}. Evidently, they are qualitatively indistinguishable. This is mainly due to the extremely small difference between $J_{xx}$ and $J_{yy}$ in the two proposed parameter sets.

In the case of Ce$_2$Sn$_2$O$_7$, we note that the magnetostriction length changes under (111) are qualitatively similar between the octupolar $(J_{xx},J_{yy},J_{zz})=(0.0104, 0.048 0.0104)$meV and $\theta=0$meV~\cite{poree2025evidence} and the dipolar parameter set $(J_{xx},J_{yy},J_{zz})=(0.045, -0.001, -0.012)$meV~\cite{yahne2024dipolar} and $\theta=0.19$ as shown in Fig.~\ref{fig:CSO Magnetostriction}(a-b) and (d-e), respectively. But, we see that $L^{(110)}$ is an order of magnitude larger in the dipolar-dominant parameter set than that of the octupolar-dominant one. Finally, we see a clear qualitative distinction in $L^{(001)}$ as we see an apparent discontinuity in the dipolar-dominant case. 

For Ce$_2$Hf$_2$O$_7$, aside from the $L^{(001)}$ results discussed in the main text, we note that the length changes between the octupolar and dipolar fitting parameters are one order of magnitudes apart under a (110) and qualitatively similar under a (111) field, as shown in Fig.~\ref{fig:CHO Magnetostriction}(a,d) and (b,e).

We want to comment on the universality of the magnetostriction signatures of Ce$_2$Hf$_2$O$_7$, since fitting parameters can differ between different experimental studies. In this work, we have used the fitting parameters of Ref.~\cite{poree2024dipolar}. Here, the drastically different behaviour between the two parameter sets mainly originates from the different $\theta$ values. This remains true as long as $J_{xz}$ of both fitting parameters are finite and on the same scale. Namely, since $\tan(2\theta)=\frac{2J_{xz}}{J_{xx}-J_{zz}}$, for the dipolar-dominant case, $J_{xx}$ is always the largest, therefore, $|J_{xx}-J_{zz}|$ is also large, resulting in an overall small $\theta$. On the other hand, for the octupolar dominant case, $|J_{xx}-J_{zz}|$ is much smaller here, resulting in a larger $\theta$. Under finite magnetic fields, the octupolar-dominant case would couple more strongly to pseudospin components $S^x_\mu$. As such, the two parameter sets would develop drastically different underlying pseudospin configurations, resulting in strikingly different magnetostriction signatures. We note that another recent study on Ce$_2$Hf$_2$O$_7$~\cite{smith2025twopeakheatcapacityaccounts} has reported slightly different fitting parameters where the best fit $\theta=0$ for both the dipolar and octupolar case with overall similar values of $(J_{xx}, J_{yy}, J_{zz})$ to that of Ref.~\cite{poree2024dipolar}. Because $\theta$ is predicted to be zero here, magnetostriction signatures between the dipolar and the octupolar-dominant case are predicted to be identical. However, it is also noted that, there, the goodness-of-fit parameter is rather shallow and a wide range of $\theta$ are in principle possible~\cite{smith2025twopeakheatcapacityaccounts}. Therefore, when considering both experimental studies holistically, we decided on reporting using Ref~\cite{poree2024dipolar}'s parameter sets.

\section{Transformation Properties of the Parton Operators \label{sec:Appendix_DOice}}
In the case where $J_{yy}$ is dominant for an octupolar spin ice, the generators of the pyrochlore space group act on the parton construction as:
\begin{widetext}
\begin{align}
T_i: & \left\{\frac{1}{2} \Phi_{\mathbf{r}_A}^{\dagger} e^{i A_{\mathbf{r}_A, \mathbf{r}_A}+\mathbf{b}_\mu} \Phi_{\mathbf{r}_A+\mathbf{b}_\mu}, \frac{1}{2} \Phi_{\mathbf{r}_A+\mathbf{b}_\mu}^{\dagger} e^{-i A_{\mathbf{r}_A, \mathbf{r}_A+\mathbf{b}_\mu}} \Phi_{\mathbf{r}_A}, E_{\mathbf{r}_A, \mathbf{r}_A+\mathbf{b}_\mu}\right\} \notag\\
& \mapsto\left\{\frac{1}{2} \Phi_{T_i\left(\mathbf{r}_A\right)}^{\dagger} e^{i A_{T_i\left(\mathbf{r}_A\right), T_i\left(\mathbf{r}_A+\mathbf{b}_\mu\right)}} \Phi_{T_i\left(\mathbf{r}_A+\mathbf{b}_\mu\right)}, \frac{1}{2} \Phi_{T_i\left(\mathbf{r}_A+\mathbf{b}_\mu\right)}^{\dagger} e^{-i A_{T_{\ell}\left(\mathbf{r}_A\right), T_{\ell}\left(\mathbf{r}_A+\mathbf{b}_\mu\right)}} \Phi_{T_i\left(\mathbf{r}_A\right)}, E_{T_i\left(\mathbf{r}_A\right), T_i\left(\mathbf{r}_A+\mathbf{b}_\mu\right)}\right\}\\
\bar{C}_6: &\left\{\frac{1}{2} \Phi_{\mathbf{r}_A}^{\dagger} e^{i A_{\mathbf{r}_A, \mathbf{r}_A}+\mathbf{b}_\mu} \Phi_{\mathbf{r}_A+\mathbf{b}_\mu}, \frac{1}{2} \Phi_{\mathbf{r}_A+\mathbf{b}_\mu}^{\dagger} e^{-i A_{\mathbf{r}_A}, \mathbf{r}_A+\mathbf{b}_\mu} \Phi_{\mathbf{r}_A}, E_{\mathbf{r}_A, \mathbf{r}_A+\mathbf{b}_\mu}\right\} \notag\\
& \mapsto\left\{\frac{1}{2} \Phi_{\bar{C}_6\left(\mathbf{r}_A\right)}^{\dagger} e^{i A_{\bar{C}_6\left(\mathbf{r}_A\right), \bar{C}_6\left(\mathbf{r}_A+\mathbf{b}_\mu\right)}} \Phi_{\bar{C}_6\left(\mathbf{r}_A+\mathbf{b}_\mu\right)}, \frac{1}{2} \Phi_{\bar{C}_6\left(\mathbf{r}_A+\mathbf{b}_\mu\right)}^{\dagger} e^{-i A_{\bar{C}_6\left(\mathbf{r}_A\right), \bar{C}_6\left(\mathbf{r}_A+\mathbf{b}_\mu\right)}} \Phi_{\bar{C}_6\left(\mathbf{r}_A\right)}, E_{\bar{C}_6\left(\mathbf{r}_A\right), \bar{C}_6\left(\mathbf{r}_A+\mathbf{b}_\mu\right)}\right\} \\
S: &\left\{\frac{1}{2} \Phi_{\mathbf{r}_A}^{\dagger} e^{i A_{\mathbf{r}_A, \mathbf{r}_A}+\mathbf{b}_\mu} \Phi_{\mathbf{r}_A+\mathbf{b}_\mu}, \frac{1}{2} \Phi_{\mathbf{r}_A+\mathbf{b}_\mu}^{\dagger} e^{-i A_{\mathbf{r}_A, \mathbf{r}_A}+\mathbf{b}_\mu} \Phi_{\mathbf{r}_A}, E_{\mathbf{r}_A, \mathbf{r}_A+\mathbf{b}_\mu}\right\} \notag\\
& \mapsto\left\{-\frac{1}{2} \Phi_{S\left(\mathbf{r}_A\right)}^{\dagger} e^{i A_{S\left(\mathbf{r}_A\right), S\left(\mathbf{r}_A+\mathbf{b}_\mu\right)}} \Phi_{S\left(\mathbf{r}_A+\mathbf{b}_\mu\right)},-\frac{1}{2} \Phi_{S\left(\mathbf{r}_A\right)}^{\dagger} e^{i A_{S\left(\mathbf{r}_A\right), S\left(\mathbf{r}_A+\mathbf{b}_\mu\right)}} \Phi_{S\left(\mathbf{r}_A+\mathbf{b}_\mu\right)},E_{S\left(\mathbf{r}_A\right), S\left(\mathbf{r}_A+\mathbf{b}_\mu\right)}\right\}.
\end{align}
\end{widetext}

We note that for the dipolar case, where $S^x\sim E$, $S^\pm=S^y\pm iS^z$, the screw operation $S$ would instead map $S: S^\pm\rightarrow S^\mp$. Under GMFT formalism, this swaps the spinon creation and annihilation operators and, therefore, will result in different PSG equations since this operation flips $\bar{A}\rightarrow -\bar{A}$. We will see in Sec.~\ref{sec:dipolar_ice} that this will result in different PSG classes.

\section{Projective Symmetry Group\label{sec:Appendix_PSG}}

\subsection{Generalities}
Due to the projective construction in GMFT, mean-field Ans\"atze that are related via some $U(1)$ gauge transformation correspond to the same physical wave function~\cite{wen2002quantum,wen2004quantum,savary2012coulombic,savary2016quantumspinliquids}. As such, mean-field Ans\"atze needs only to be invariant under $\mathcal{G}_\mathcal{O}\mathcal{O}$, where $\mathcal{O}$ is some symmetry operation and $\mathcal{G}_\mathcal{O}:|\Psi_{\mathbf{r}_\alpha}\rangle \rightarrow e^{i\phi_\mathcal{O}(\mathbf{r}_\alpha)}|\Psi_{\mathbf{r}_\alpha}\rangle$ is some $U(1)$ transformation associated with the symmetry operation. 

The projective representations of the space group generators $\mathcal{G}_\mathcal{O}\mathcal{O}$ must follow the same underlying algebraic relation. The PSG classification of all fully symmetric states is therefore found by imposing such algebraic constraints. However, this is not enough to unambiguously find a representation of the fully symmetric states due to the fact that there is a gauge freedom in $\mathcal{G}_\mathcal{O}$. A PSG element $\mathcal{G}_\mathcal{O} \mathcal{O}\in \text{PSG}$ transforms under some general gauge transformation $\mathcal{G}:\ket{\Psi_{\mathbf{r}_\alpha}}=e^{i\phi(\mathbf{r}_\alpha)}\ket{\Psi_{\mathbf{r}_\alpha}}$ as
\begin{align}
   \mathcal{G}_\mathcal{O} \mathcal{O} \rightarrow \mathcal{G}\mathcal{G}_\mathcal{O} \mathcal{O} \mathcal{G}^{-1} &= \mathcal{G} \mathcal{G}_\mathcal{O} \mathcal{O} \mathcal{G}^{-1} \mathcal{O}^{-1}\mathcal{O} \\
   &= \mathcal{G}\mathcal{G}_\mathcal{O} \mathcal{G}^{-1}[\mathcal{O}^{-1}(\mathbf{r}_\alpha)]\mathcal{O},
\end{align}
where we used the relation
\begin{equation}
    \mathcal{O} \mathcal{G}\mathcal{O}^{-1} = \mathcal{G}[\mathcal{O}^{-1}(\mathbf{r}_\alpha)].
\end{equation}
Here $\mathcal{G}[\mathcal{O}^{-1}(\mathbf{r}_\alpha)]:\ket{\Psi_{\mathbf{r}_\alpha}}\rightarrow e^{i\phi(\mathcal{O}^{-1}(\mathbf{r}_\alpha))}\ket{\Psi_{\mathbf{r}_\alpha}}$
As such, the PSG phase transforms as
\begin{equation}
    \phi_\mathcal{O}(\mathbf{r}_\alpha) \rightarrow \phi_\mathcal{O}(\mathbf{r}_\alpha) + \phi(\mathbf{r}_\alpha) - \phi(\mathcal{O}^{-1}(\mathbf{r}_\alpha)).
\end{equation}
To obtain an unambiguous representation of the PSG class, we must fix the gauge freedoms. If we assume spatially isotropic and translationally invariant phase factors, there are 6 distinct gauge transformations corresponding to 2 diamond sublattices and 3 directions.
\begin{equation}
    \phi_{i,\beta}(\mathbf{r}_\alpha) =  \psi_{i,\beta} r_i \delta_{\alpha,\beta} \label{eq:gauge1},
\end{equation}
where $\psi_{i,\beta}$ is some $U(1)$ element. Furthermore, one can apply a sublattice-dependent gauge transformation:
\begin{equation}
    \bar{\phi}(\mathbf{r}_\alpha) =  \bar{\psi}_{\beta} \delta_{\alpha,\beta} \label{eq:gauge2},
\end{equation}
$\psi_{i,\beta}$, $\bar{\psi}_\beta \in [0, 2\pi)$. In total, we have 8 gauge degrees of freedom that we will use in the following sections to completely fix the phase factors along with site-independent $U(1)$ gauge transformation associated with each operation.

\subsection{PSG for Octupolar Spin Ice under Magnetic Fields}

Under finite magnetic fields, the point group symmetry of the pyrochlore lattice is lowered depending on the imposed magnetic field direction. As such, we only need to calculate the PSG solutions for the generators of the remaining point group. 

Solutions of PSG follow exactly the same as that of Ref.~\cite{zhou2024magnetic}. Here, we summarize the necessary results. Under a (110) magnetic field, the generators of the remaining point group is a reflection $\sigma=S\bar{C}_6^3$ and inversion $I=\bar{C}_6^3$. The resulting PSG solution is:
\begin{subequations}
\begin{align}
&\phi_{T_1}(\mathbf{r}_\alpha) = 0\\
&\phi_{T_2}(\mathbf{r}_\alpha) = n_1\pi r_1\\
&\phi_{T_3}(\mathbf{r}_\alpha) = n_2\pi (r_1+r_2)\\
&\phi_\sigma(\mathbf{r}_\alpha) = n_2\pi (r_1+r_2)(r_1+r_2+1)/2\\
&\phi_I(\mathbf{r}_\alpha) = \psi_{IT_1}r_1 + \psi_{IT_2}r_2 + \psi_I \delta_{\alpha, B}.
\end{align}
\end{subequations}
Here $n_1,n_2\in\mathbb{Z}_2$; $\psi_{IT_1}, \psi_{IT_2}, \psi_{I}\in U(1)$

On the other hand for a (111) field, the only remaining generator of the point group is $\bar{C}_6$. Therefore, we get the following PSG solution:
\begin{subequations}
\begin{align}
&\phi_{T_1}(\mathbf{r}_\alpha) = 0\\
&\phi_{T_2}(\mathbf{r}_\alpha) = n_1\pi r_1\\
&\phi_{T_3}(\mathbf{r}_\alpha) = n_1\pi (r_1+r_2)\\
&\phi_{C}(\mathbf{r}_\alpha) = n_1\pi r_1(r_2+r_3) + \frac{1}{3}\psi_{C_6}\delta_{\alpha,B},
\end{align}
\end{subequations}
where $n_1\in\mathbb{Z}_2$ and $\psi_{C_6}\in U(1)$. 

Finally, for the (001) case, here the remaining generators are $\bar{C}_4=\bar{C}_6^2S\bar{C}_6^{-1}$ and inversion. The resulting PSG solution is: 
\begin{subequations}
\begin{align}
&\phi_{T_1}(\mathbf{r}_\alpha) = 0\\
&\phi_{T_2}(\mathbf{r}_\alpha) = n_1\pi r_1\\
&\phi_{T_3}(\mathbf{r}_\alpha) = n_1\pi (r_1+r_2)\\
&\phi_{\bar{C}_4}(\mathbf{r}_\alpha) = \frac{n_1\pi}{2}(-r_1(1+r_1)+r_3(1+r_3))\notag \\
\quad\quad&+n_1\pi r_1 r_2+n_1\pi(r_1+r_3)\delta_{\alpha,B}+\psi_{\bar{C}_4}/4\\
&\phi_I(\mathbf{r}_\alpha) = 0
\end{align}
\end{subequations}
Here, again, $n_1\in\mathbb{Z}_2$, $\psi_{\bar{C}_4}\in U(1)$.

\subsection{PSG for Dipolar Spin Ice under Magnetic Fields\label{sec:dipolar_ice}}

In the case where $J_{xx}$ is dominant, the screw operation $S$ will instead map creation operators to annihilation operators and vice versa. Therefore, the PSG will differ for symmetry group with generators containing the screw operation $S$. These operators are $\sigma$ under $[110]$ fields and $\bar{C}_4$ under $[001]$ fields. Therefore, we expect different PSG results for these fields. After going through similar derivations, it turns out that under a $[110]$ field, the PSG solution is:
\begin{subequations}
\begin{align}
\phi_{T_1}(\mathbf{r}_\alpha) &= 0\\
\phi_{T_2}(\mathbf{r}_\alpha) &= n_1\pi r_1\\
\phi_{T_3}(\mathbf{r}_\alpha) &= n_2\pi (r_1+r_2)\\
\phi_\sigma(\mathbf{r}_\alpha) &= n_2\pi (r_1+r_2)(r_1+r_2+1)/2\notag\\
&\quad -\psi_{\sigma T_1}r_1-\psi_{\sigma T_2}r_2\label{eq:sxx1}\\
\phi_I(\mathbf{r}_\alpha) &= n_I\pi(r_1+r_2)+ \psi_I/2\label{eq:sxx2}.
\end{align}
\end{subequations}
where $n_1,n_2,n_I\in\mathbb{Z}_2$. The PSG solutions under a $[001]$ field is:
\begin{subequations}
\begin{align}
&\phi_{T_1}(\mathbf{}{r}_\alpha) = 0\\
&\phi_{T_2}(\mathbf{r}_\alpha) = n_1\pi r_1\\
&\phi_{T_3}(\mathbf{r}_\alpha) = n_1\pi (r_1+r_2)\\
&\phi_{\bar{C}_4}(\mathbf{r}_\alpha) = \frac{n_1\pi}{2}(-r_1(1+r_1)+r_3(1+r_3))\notag \\
&\qquad\quad+n_1\pi r_1 r_2+(n_1\pi(r_1+r_3))\delta_{\alpha,B}+\psi_{\bar{C}_4}\label{eq:sxx3}\\
&\phi_I(\mathbf{r}_\alpha) = n_I\pi.\label{eq:sxx4}
\end{align}
\end{subequations}
where $n_1,n_I\in\mathbb{Z}_2$, $\psi_{\bar{C}_4}\in U(1)$. As such, the PSG solution for the dipolar case differs from that of the octupolar case. 
Despite this difference, both result in the same mean-field gauge configuration $\bar{A}_{\mathbf{r}_\alpha, \mathbf{r}_\alpha+\mathbf{b}_\mu}$ as in Eqs.~\eqref{eq:A110} and~\eqref{eq:A001}. However, the mean-field parameter $\chi$, $\xi$ are indeed different, as we will demonstrate in Sec.~\ref{sec:Appendix_MFP}.

\subsection{Mean Gauge Configuration}
After getting all the PSG solutions, we can now find the fully symmetric mean-field solution by looking at how they are related under the remaining symmetry operations. Indeed, the MF Hamiltonian has to be invariant under the projective transformations $\mathcal{G}_\mathcal{O}\mathcal{O}$. This determines the 
corresponding mean-field Ans\"atze $\mathcal{G}_{\bar{A}}(\mathbf{r}_A,\mathbf{r}_A+b_\mu)= e^{iA_{\mathbf{r}_A,\mathbf{r}_A+b_\mu}}$. More specifically, invariance of the MF Hamiltonian under $\mathcal{G}_\mathcal{O}\mathcal{O}$ implies
\begin{equation}
\begin{aligned}
    \mathcal{G}_{\bar{A}} &(O(\mathbf{r}_A), O(\mathbf{r}_A + \mathbf{b}_\mu)) \\
    &= \mathcal{G}_O^\dagger(O(\mathbf{r}_A)) \mathcal{G}_{\bar{A}} (\mathbf{r}_A, \mathbf{r}_A + \mathbf{b}_\mu) \mathcal{G}_O(O(\mathbf{r}_A)).\label{eq:gaugeconfigrelateion}
\end{aligned}
\end{equation}
From these requirements, the gauge field configuration on the entire lattice can be determined for a given PSG class. To do so, the value of the gauge field background is arbitrarily fixed on representative bonds that are not related to each other but are related by all other bonds on the lattice by symmetry operations. In the presence of translation symmetry, this inequivalent set can be taken to contain at most the for bonds of the unit cell at the origin $\bar{A}_{\mathbf{0}_A, \mathbf{0}_A+\mathbf{b}_\mu}:=\bar{A}_\mu$. In many cases, these bonds might be related by symmetry operations. We compute the mean field gauge configuration for the octupolar-dominant case, although it turns out that the dipolar-dominant case has the exact same relations.

For the [110] field, the bonds $\bar{A}_{\mathbf{0}_A, \mathbf{0}_A+\mathbf{b}_0}$, $\bar{A}_{\mathbf{0}_A, \mathbf{0}_A+\mathbf{b}_1}$, and $\bar{A}_{\mathbf{0}_A, \mathbf{0}_A+\mathbf{b}_2}$ cannot be related by the remaining symmetry transformations. The mean-field configuration for a specific PSG class is then specified by three parameters $\bar{A}_0$, $\bar{A}_1$, and $\bar{A}_2$:
\begin{subequations}\label{eq:A110}
\begin{align}
& A\left[\left(r_1, r_2, r_3\right)_A,\left(r_1, r_2, r_3\right)_B\right]=\bar{A}_0  \\
& A\left[\left(r_1, r_2, r_3\right)_A,\left(r_1+1, r_2, r_3\right)_B\right]=\bar{A}_1+n_1 \pi r_2+n_2\pi r_3   \\
& A\left[\left(r_1, r_2, r_3\right)_A,\left(r_1, r_2+1, r_3\right)_B\right]=\bar{A}_2+n_2 \pi r_3  \\
& A\left[\left(r_1, r_2, r_3\right)_A,\left(r_1, r_2, r_3+1\right)_B\right]=\bar{A}_0.
\end{align}
\end{subequations}

Similarly, we can only mix the bonds $\bar{A}_{\mathbf{0}_A, \mathbf{0}_A+\mathbf{b}_1}$, $\bar{A}_{\mathbf{0}_A, \mathbf{0}_A+\mathbf{b}_2}$ and $\bar{A}_{\mathbf{0}_A, \mathbf{0}_A+\mathbf{b}_3}$ by symmetry operations in the presence of a [111] field. The gauge field background configuration of the associated PSG class then depends on two parameters $\bar{A}_0$ and $\bar{A}_1$:
\begin{subequations}\label{eq:A111}
\begin{align}
& A\left[\left(r_1, r_2, r_3\right)_A,\left(r_1, r_2, r_3\right)_B\right]=\bar{A}_0 \\
& A\left[\left(r_1, r_2, r_3\right)_A,\left(r_1+1, r_2, r_3\right)_B\right]=\bar{A}_1+n_1 \pi\left(r_2+r_3\right) \\
& A\left[\left(r_1, r_2, r_3\right)_A,\left(r_1, r_2+1, r_3\right)_B\right]=\bar{A}_1+n_1 \pi r_3 \\
& A\left[\left(r_1, r_2, r_3\right)_A,\left(r_1, r_2, r_3+1\right)_B\right]=\bar{A}_1
\end{align}
\end{subequations}

Finally, for the $[001]$ field, $\bar{C}_4$ can indeed relate all 4 sites. Therefore, the gauge field configuration would depend on the parameter $\bar{A}_{0}$:
\begin{subequations}\label{eq:A001}
\begin{align}
& A\left[\left(r_1, r_2, r_3\right)_A,\left(r_1, r_2, r_3\right)_B\right]=\bar{A}_0 \\
& A\left[\left(r_1, r_2, r_3\right)_A,\left(r_1+1, r_2, r_3\right)_B\right]=\bar{A}_0+n_1 \pi (r_2+r_3) \\
& A\left[\left(r_1, r_2, r_3\right)_A,\left(r_1, r_2+1, r_3\right)_B\right]=\bar{A}_0+ n_1 \pi r_3 \\
& A\left[\left(r_1, r_2, r_3\right)_A,\left(r_1, r_2, r_3+1\right)_B\right]=\bar{A}_0
\end{align}
\end{subequations}

\subsection{Mean Field Parameters\label{sec:Appendix_MFP}}

We can use the same relations to relate the other mean-field parameters. Let us define $\xi_{\mathbf{0}_A, \mathbf{0}_A+\mathbf{b}_\mu}:=\xi_{\mu}$ and $\chi_{\mathbf{0}_\alpha+\mathbf{b}_\mu, \mathbf{0}_\alpha+\mathbf{b}_\nu}:=\chi_{\mu\nu}^\beta$, where $\beta=A \textrm{  if  } \alpha = B$ and vice versa. Notice that by construction $\xi_{\mathbf{0}_A, \mathbf{0}_A+\mathbf{b}_\mu} = \xi_{\mathbf{0}_A+\mathbf{b}_\mu, \mathbf{0}_A}^*$.

Under a (110) field, the mean-field parameters relate as:
\begin{subequations}\label{eq:xi110}
\begin{align}
& \xi\left[\left(r_1, r_2, r_3\right)_A,\left(r_1, r_2, r_3\right)_B\right]=\xi_{0}  \\
& \xi\left[\left(r_1, r_2, r_3\right)_A,\left(r_1+1, r_2, r_3\right)_B\right]=\xi_{1} e^{i(n_1\pi r_2 +n_2\pi r_3)}  \\
& \xi\left[\left(r_1, r_2, r_3\right)_A,\left(r_1, r_2+1, r_3\right)_B\right]=\xi_{2} e^{in_2\pi r_3} \\
& \xi\left[\left(r_1, r_2, r_3\right)_A,\left(r_1, r_2, r_3+1\right)_B\right]=\xi_{0}.
\end{align}
\end{subequations}
Here, we can see that since Inversion is still present under a (110) field, we can apply $I$ on $\xi_{0}=\xi_{\mathbf{0}_A, \mathbf{0}_B} = \xi_{\mathbf{0}_B, \mathbf{0}_A} = \xi_{0}^*$. Therefore, $\xi_{0}\in\mathbb{R}$. Similarly, $\xi_{1}=\xi_{\mathbf{0}_A, \mathbf{0}_A+\mathbf{b}_1} = \xi_{\mathbf{0}_A-\mathbf{b}_1, \mathbf{0}_A} = \xi_{1}^*$, since $\xi_1$ does not depend on $\mathbf{r}_1$. The same can be said about $\xi_{2}$. Therefore, $\xi_1$ and $\xi_2$ are both real. On the other hand, mean-field parameter $\chi$ are related in the following ways:
\begin{subequations}\label{eq:chi110}
\begin{align}
& \chi\left[\left(r_1, r_2, r_3\right)_B,\left(r_1+1, r_2, r_3\right)_B\right]=\chi_{01}e^{i\pi(n_1 r_2 +in_2 r_3)}  \\
& \chi\left[\left(r_1, r_2, r_3\right)_B,\left(r_1, r_2+1, r_3\right)_B\right]=\chi_{02} e^{in_2\pi r_3} \\
& \chi\left[\left(r_1, r_2, r_3\right)_B,\left(r_1, r_2, r_3+1\right)_B\right]=\chi_{03}  \\
& \chi\left[\left(r_1+1, r_2, r_3\right)_B,\left(r_1, r_2+1, r_3\right)_B\right]=\chi_{12} e^{in_1\pi r_2}  \\
& \chi\left[\left(r_1+1, r_2, r_3\right)_B,\left(r_1, r_2, r_3+1\right)_B\right]=\chi_{01} e^{i\pi(n_1r_2+ n_2 (r_3+1)}  \\
& \chi\left[\left(r_1, r_2+1, r_3\right)_A,\left(r_1, r_2, r_3+1\right)_B\right]=\chi_{12} e^{in_2 r_3}
\end{align}
\end{subequations}

And that $\chi_{0\mu}^A = e^{-i\psi_I}\chi_{0\mu}^B$ for $\mu\in\{0,1,2,3\}$; $\chi_{12}^A = \chi_{12}^B e^{-\psi_I+\psi_{IT_1}-\psi_{IT_2}}$; $\chi_{13}^A = \chi_{13}^B e^{-\psi_I+\psi_{IT_1}}$; $\chi_{23}^A = \chi_{23}^B e^{-\psi_I+\psi_{IT_2}}$. 

Just like the mean-field configuration $\bar{A}_\mu$, due to the reduced symmetry introduced by the magnetic fields, not all mean-field parameters can be related to each other. As such, we are left with 3 independent $\xi_\mu$ parameters and 4 independent $\chi_{\mu\nu}$ parameters to solve. Furthermore, the complex phase factor $\psi_{I}$, $\psi_{IT_1}$, $\psi_{IT_2}$ actually comes into the mean-field calculation now and needs to be determined self-consistently.

Under a (111) field, we have that the three types of mean-field parameters follow that:
\begin{subequations}\label{eq:xi111}
\begin{align}
& \xi\left[\left(r_1, r_2, r_3\right)_B,\left(r_1, r_2, r_3\right)_B\right]=\xi_{0}  \\
& \xi\left[\left(r_1, r_2, r_3\right)_A,\left(r_1+1, r_2, r_3\right)_B\right]=\xi_{1} e^{in_1\pi (r_2 +r_3)}  \\
& \xi\left[\left(r_1, r_2, r_3\right)_A,\left(r_1, r_2+1, r_3\right)_B\right]=\xi_{1} e^{in_1\pi r_3} \\
& \xi\left[\left(r_1, r_2, r_3\right)_A,\left(r_1, r_2, r_3+1\right)_B\right]=\xi_{1}.
\end{align}
\end{subequations}
Here, $\xi_0$ and $\xi_1$ are real by the same logic.
\begin{subequations}\label{eq:chi111}
\begin{align}
& \chi\left[\left(r_1, r_2, r_3\right)_B,\left(r_1+1, r_2, r_3\right)_B\right]=\chi_{01}e^{in_1 (r_2 + r_3)}  \\
& \chi\left[\left(r_1, r_2, r_3\right)_B,\left(r_1, r_2+1, r_3\right)_B\right]=\chi_{01} e^{i(n_1\pi r_3-4\psi_{C_6}/3)} \\
& \chi\left[\left(r_1, r_2, r_3\right)_B,\left(r_1, r_2, r_3+1\right)_B\right]=\chi_{01}e^{-i2\psi_{C_6}/3}  \\
& \chi\left[\left(r_1+1, r_2, r_3\right)_B,\left(r_1, r_2+1, r_3\right)_B\right]=\chi_{23} e^{in_1\pi r_2-i2\psi_{C_6}/3}  \\
& \chi\left[\left(r_1+1, r_2, r_3\right)_B,\left(r_1, r_2, r_3+1\right)_B\right]=\chi_{23} e^{in_1\pi(r_2+r_3)-i4\psi_{C_6}/3}  \\
& \chi\left[\left(r_1, r_2+1, r_3\right)_A,\left(r_1, r_2, r_3+1\right)_B\right]=\chi_{23} e^{in_1 r_3}
\end{align}
\end{subequations}
And the sublattice A mean-field parameters are related by the complex phase factor $\psi_{C_6}$: $\chi_{\mu\nu}^A = \chi_{\mu\nu}^Be^{-\psi_{C_6}}$ for $\mu,\nu\in\mathbb{Z}_4$.

Finally, under a (001) field, the mean-field parameters follow:
\begin{subequations}\label{eq:xi001}
\begin{align}
& \xi\left[\left(r_1, r_2, r_3\right)_B,\left(r_1, r_2, r_3\right)_B\right]=\xi_{0}  \\
& \xi\left[\left(r_1, r_2, r_3\right)_A,\left(r_1+1, r_2, r_3\right)_B\right]=\xi_{0} e^{in_1\pi (r_2 +r_3)}  \\
& \xi\left[\left(r_1, r_2, r_3\right)_A,\left(r_1, r_2+1, r_3\right)_B\right]=\xi_{0} e^{in_1\pi r_3} \\
& \xi\left[\left(r_1, r_2, r_3\right)_A,\left(r_1, r_2, r_3+1\right)_B\right]=\xi_{0}.
\end{align}
\end{subequations}
where $\xi_0 \in \mathbb{R}$.
\begin{subequations}\label{eq:chi001}
\begin{align}
& \chi\left[\left(r_1, r_2, r_3\right)_B,\left(r_1+1, r_2, r_3\right)_B\right]=\chi_{01}e^{in_1 (r_2 + r_3)}  \\
& \chi\left[\left(r_1, r_2, r_3\right)_B,\left(r_1, r_2+1, r_3\right)_B\right]=\chi_{01} e^{i(n_1\pi r_3-\psi_{\bar{C}_4}/2)} \\
& \chi\left[\left(r_1, r_2, r_3\right)_B,\left(r_1, r_2, r_3+1\right)_B\right]=\chi_{03}  \\
& \chi\left[\left(r_1+1, r_2, r_3\right)_B,\left(r_1, r_2+1, r_3\right)_B\right]=\chi_{03} e^{in_1\pi r_2-i\psi_{\bar{C}_4}/2}  \\
& \chi\left[\left(r_1+1, r_2, r_3\right)_B,\left(r_1, r_2, r_3+1\right)_B\right]=\chi_{01} e^{in_1\pi(r_2+r_3)-i\psi_{\bar{C}_4}/2}  \\
& \chi\left[\left(r_1, r_2+1, r_3\right)_A,\left(r_1, r_2, r_3+1\right)_B\right]=\chi_{01} e^{in_1 r_3-i\psi_{\bar{C}_4}}
\end{align}
\end{subequations}
Here the sublattice A mean-field parameters $\chi_{\mu\nu}^A=\chi_{\mu\nu}^B$ since $\phi_I(\mathbf{r}_\alpha)=0$. 

Similarly, we can do the same for dipolar-dominant QSI. Here we note that since the PSG solutions under a (111) field for dipolar and octupolar-dominant cases are the same, we only need to calculate for the (110) and (001) fields. Under a (110) field, the mean-field parameters relate as:
\begin{subequations}\label{eq:xi110}
\begin{align}
& \xi\left[\left(r_1, r_2, r_3\right)_A,\left(r_1, r_2, r_3\right)_B\right]=\xi_{0}  \\
& \xi\left[\left(r_1, r_2, r_3\right)_A,\left(r_1+1, r_2, r_3\right)_B\right]=\xi_{1} e^{i(n_1\pi r_2 +n_2\pi r_3)}  \\
& \xi\left[\left(r_1, r_2, r_3\right)_A,\left(r_1, r_2+1, r_3\right)_B\right]=\xi_{2} e^{in_2\pi r_3} \\
& \xi\left[\left(r_1, r_2, r_3\right)_A,\left(r_1, r_2, r_3+1\right)_B\right]=\xi_{0}.
\end{align}
\end{subequations}
Here, $\xi_0,\xi_1,\xi_2\in\mathbb{R}$.
\begin{subequations}\label{eq:chi110_dipolar}
\begin{align}
& \chi\left[\left(r_1, r_2, r_3\right)_B,\left(r_1+1, r_2, r_3\right)_B\right]=\chi_{01}e^{i\pi(n_1 r_2 +in_2 r_3)}  \\
& \chi\left[\left(r_1, r_2, r_3\right)_B,\left(r_1, r_2+1, r_3\right)_B\right]=\chi_{02} e^{in_2\pi r_3} \\
& \chi\left[\left(r_1, r_2, r_3\right)_B,\left(r_1, r_2, r_3+1\right)_B\right]=\chi_{03}  \\
& \chi\left[\left(r_1+1, r_2, r_3\right)_B,\left(r_1, r_2+1, r_3\right)_B\right]=\chi_{12} e^{in_1\pi r_2}  \\
& \chi\left[\left(r_1+1, r_2, r_3\right)_B,\left(r_1, r_2, r_3+1\right)_B\right]=\chi_{01} e^{i\pi(n_1r_2+ n_2 (r_3+1)+\psi_{\sigma T_1}}  \\
& \chi\left[\left(r_1, r_2+1, r_3\right)_A,\left(r_1, r_2, r_3+1\right)_B\right]=\chi_{12} e^{i(n_2 r_3+\psi_{\sigma T_1}+\psi_{\sigma T_2})}
\end{align}
\end{subequations}

$\chi_{00}^A = \chi_{00}^Be^{-i\psi_I}$; ;$\chi_{01}^A = \chi_{01}^Be^{-i\psi_I+in_I}$; $\chi_{02}^A = \chi_{02}^Be^{-i\psi_I+in_I}$; $\chi_{03}^A = \chi_{03}^Be^{-i\psi_I}$; $\chi_{12}^A = \chi_{12}^Be^{-i\psi_I}$ $\chi_{13}^A = \chi_{13}^Be^{i\psi_{\sigma T_1}-i\psi_{I}+in_2\pi+in_I\pi}$; $\chi_{23}^A = \chi_{23}^Be^{i\psi_{\sigma T_1}+i\psi_{\sigma T_2}-in_I\pi -i\psi_{I}}$;

Finally, under a (001) field, the mean-field parameters follow:
\begin{subequations}\label{eq:xi001}
\begin{align}
& \xi\left[\left(r_1, r_2, r_3\right)_B,\left(r_1, r_2, r_3\right)_B\right]=\xi_{0}  \\
& \xi\left[\left(r_1, r_2, r_3\right)_A,\left(r_1+1, r_2, r_3\right)_B\right]=\xi_{0} e^{in_1\pi (r_2 +r_3)}  \\
& \xi\left[\left(r_1, r_2, r_3\right)_A,\left(r_1, r_2+1, r_3\right)_B\right]=\xi_{0} e^{in_1\pi r_3} \\
& \xi\left[\left(r_1, r_2, r_3\right)_A,\left(r_1, r_2, r_3+1\right)_B\right]=\xi_{0}.
\end{align}
\end{subequations}
where $\xi_0 \in \mathbb{R}$.
\begin{subequations}\label{eq:chi001}
\begin{align}
& \chi\left[\left(r_1, r_2, r_3\right)_B,\left(r_1+1, r_2, r_3\right)_B\right]=\chi_{01}e^{in_1 (r_2 + r_3)}  \\
& \chi\left[\left(r_1, r_2, r_3\right)_B,\left(r_1, r_2+1, r_3\right)_B\right]=\chi_{01} e^{i(n_1\pi r_3-3\psi_{\bar{C}_4})} \\
& \chi\left[\left(r_1, r_2, r_3\right)_B,\left(r_1, r_2, r_3+1\right)_B\right]=\chi_{03}  \\
& \chi\left[\left(r_1+1, r_2, r_3\right)_B,\left(r_1, r_2+1, r_3\right)_B\right]=\chi_{03} e^{in_1\pi r_2-i\psi_{\bar{C}_4}}  \\
& \chi\left[\left(r_1+1, r_2, r_3\right)_B,\left(r_1, r_2, r_3+1\right)_B\right]=\chi_{01} e^{in_1\pi(r_2+r_3)-i\psi_{\bar{C}_4}}  \\
& \chi\left[\left(r_1, r_2+1, r_3\right)_A,\left(r_1, r_2, r_3+1\right)_B\right]=\chi_{01} e^{in_1 r_3-i2\psi_{\bar{C}_4}}
\end{align}
\end{subequations}
Here the sublattice A mean-field parameters $\chi_{\mu\nu}^A=\chi_{\mu\nu}^B$ since $\phi_I(\mathbf{r}_\alpha)=0$.

We have now successfully categorized all the mean field parameters under different magnetic field directions. Now, we are equipped to solve the mean-field Hamiltonian in Eq.~\ref{eq:H_GMFT} self-consistently by evaluating the respective mean field while respecting the relations enumerated above. Furthermore, all complex phase factors $\psi$ also need to be solved self-consistently. Clearly, the number of free parameters makes the convergence of the problem computationally costly or inaccurate if fell into some local minimum. However, a phase diagram is indeed attainable with sufficient resolution of the parameter space as well as enough computational resources.

\bibliography{ref}